\documentclass[aps,pre,epsf,reprint,onecolumn,superscriptaddress,showpacs,amsmath,amssymb,amsfonts,nofootinbib,8pt]{revtex4-1}
\usepackage{epsfig}
\usepackage{amsmath}
\usepackage{mathtools}
\usepackage{color}
\usepackage{graphicx}
\usepackage{dcolumn}
\usepackage{bm}
\usepackage{braket}
\usepackage[T1]{fontenc}
\begin{document}

\title{Probing Ferromagnetic Order in Few-Fermion\\ Correlated Spin-Flip Dynamics}

\author{G.M. Koutentakis}
\affiliation{Zentrum f\"{u}r Optische Quantentechnologien,
Universit\"{a}t Hamburg, Luruper Chaussee 149, 22761 Hamburg,
Germany}\affiliation{The Hamburg Centre for Ultrafast Imaging,
Universit\"{a}t Hamburg, Luruper Chaussee 149, 22761 Hamburg,
Germany}
\author{S.I. Mistakidis}
\affiliation{Zentrum f\"{u}r Optische Quantentechnologien,
Universit\"{a}t Hamburg, Luruper Chaussee 149, 22761 Hamburg,
Germany}
\author{P. Schmelcher}
\affiliation{Zentrum f\"{u}r Optische Quantentechnologien,
Universit\"{a}t Hamburg, Luruper Chaussee 149, 22761 Hamburg,
Germany} \affiliation{The Hamburg Centre for Ultrafast Imaging,
Universit\"{a}t Hamburg, Luruper Chaussee 149, 22761 Hamburg,
Germany}

\date{\today}

\begin{abstract} 
    We unravel the dynamical stability of a fully polarized one-dimensional ultracold few-fermion
    spin-$1/2$ gas subjected to inhomogeneous driving of the itinerant spins. Despite the unstable character of the total
    spin-polarization the existence of an
    interaction regime is demonstrated where the spin-correlations lead to almost 
    maximally aligned spins throughout the dynamics. The resulting ferromagnetic order emerges from the build up of superpositions of
    states of maximal total spin.  They comprise a decaying  spin-polarization and
    a dynamical evolution towards an almost completely unpolarized NOON-like state.
    Via single-shot simulations we demonstrate that our theoretical predictions can be detected in state-of-the-art ultracold experiments.
\end{abstract}

\maketitle

\section{INTRODUCTION}
Magnetism constitutes a principal feature of a large
class of materials and represents a macroscopic phenomenon of quantum origin
\cite{Vollhardt,Brando,Sachdev}.  In conductors the magnetic properties of the
delocalized (itinerant) electrons are qualitatively understood in terms of the
Stoner instability \cite{Stoner}.  To verify and emulate the latter mechanism
ultracold fermionic ensembles have been employed
\cite{Ketterle1,Ketterle2,Pekker}. However, the nature of the interparticle interaction
exhibited in three-dimensional ultracold gases hindered the study of itinerant
ferromagnetism as the repulsive Fermi gas is metastable due to bosonic Feshbach
molecule formation \cite{Chin}.  Utilizing fast interaction quenches, it has
been shown that no ferromagnetic phase can be achieved as the decay into
molecules is faster than the formation of ferromagnetic domains
\cite{Ketterle2,Pekker}. Instead, recent pump-probe experiments \cite{LENS-Ketterle} indicate that 
the formation rate of ferromagetic domains with a size comparable to the interatomic separation is larger than the corresponding molecular decay rate.
Furthermore, ferromagnetic properties have been
observed indirectly either by the spectroscopic study of strongly
particle-imbalanced \cite{Grimm,Scazza} (supplemented by \cite{Li_objection})
and particle-balanced \cite{LENS-Ketterle} two-component Fermi mixtures or by
employing a binary Fermi gas prepared in a magnetic domain wall structure
\cite{Valtolina}. The latter experimental evidence poses the question whether
stable ferromagnetism can be observed in the absence of molecule formation.
 
A controllable setting that can shed light on such inquiries is the
experimentally accessible few-fermion quasi-one dimensional gas
\cite{Heidelberg}. Owing to its one-dimensional (1D) character, a shallow two-body
bound state for effectively repulsive interactions is absent and thus
the molecule formation is suppressed. Moreover, the experimental
\cite{BrouzosJochim,ZurnAntiferro} and theoretical
\cite{Blume,BrouzosFer,Buignion,Lewenstein,Zinner} study of the magnetic
properties of few-fermion systems has led to the insight that for near zero
and infinite interactions there is an approximate mapping of the 1D spin-$1/2$ Fermi gas to an effective spin-chain
model \cite{Levinsen,Deuretzbacher,Zinnerchain,Yangchain,Cuitrns,Cui_ref1,AMR1}. Most importantly,
these spinor systems possess experimentally accessible eigenstates of ferromagnetic nature\footnote{For our purposes, a state is defined as ferromagnetic when it is characterized by
maximal spin alignment and polarization.}, namely
the interaction-independent spin-polarized states. Consequently, the study of
the dynamical stability of these ferromagnetic states
is essential for our understanding of the magnetic properties of 1D systems.
The study of the mechanisms emanating in 1D systems might in turn provide important 
insights for higher dimensional settings.

We study the dynamical stability of the fully polarized 1D parabolically-confined few-fermion
spin-$1/2$ gas under the effect of inhomogeneous Rabi coupling of the spin-states.  This coupling
scheme introduces a spatially dependent spin-flip transition amplitude and thus probes the stability
of the initial state by breaking the spin-symmetries of the unperturbed system (see below). An argumentation based on the
Hartree-Fock (HF) framework of the Stoner model testified within the
time-dependent Hartree-Fock (TDHF) \cite{Grochowski} showed that the
spin-polarization of the Fermi gas is stable for interparticle repulsions that exceed the kinetic
and spin-flip contributions \cite{Salasnich,Salasnich1}. Inspecting the correlated
spin-flip dynamics within the
latter interaction regime (where TDHF predicts stable ferromagnetism) we reveal that the many-body (MB) spin-spin
correlator exhibits ferromagnetic spin-spin correlations throughout the dynamics.
Moreover and in contrast to the TDHF results the MB state of the ferromagnetically correlated gas shows an
unstable polarization fluctuating between fully polarized and almost completely unpolarized. This
outcome cannot be retrieved within the HF description and exposes the crucial role of
correlations in the magnetic properties of spin-$1/2$ fermions even away from the
strongly interacting regime. We show that the decay of the polarization and the emerging correlated
spin-order can be understood by
generalizing the spin-chain model of Ref. \cite{AMR1}. The coupling of the initial state to lower
spin-$S$ values is found to cause the dephasing of the collective Larmor precession of the spins.
This dephasing dynamically leads the system to an almost equal superposition of the two
ferromagnetic fully-polarized states of opposite spin-orientation 
i.e. a NOON-like state \cite{NOON1,NOON2} with zero total polarization. For weaker and stronger interactions lying outside the
above-mentioned ferromagnetically ordered regime the system undergoes a demagnetization dynamics which
is identified and characterized. Our results generalize to other particle numbers
within the few-body regime. The employed setup can be implemented in state-of-the-art ${}^{40}$K
experiments and the corresponding findings can be probed by fluorescence imaging techniques.
Additionally, we showcase that our findings can be generalized to a broader class of dynamical scenarios characterized by 
different initial states and Rabi-coupling potentials. We
explicitly demonstrate the robustness of our results against common noise sources exhibited in such
experiments by performing simulations of single-shot images.

The presentation of our results proceeds as follows. In section II we discuss the setup and the
basic observables used for the interpretation of the spin-dynamics.  In section III we present our
results for the correlated spin-flip dynamics employing the Multi-Configuration Time-Dependent
Method for Fermions (MCTDHF) \cite{Alon,Axel,MLX,Jenny1,Jenny2,Fpolarons} and interpret them in
terms of two approximate methods for the case $N=3$. Section IV provides a generalization of our
findings to the case of $N=5$ fermions. In section V we also demonstrate that the observed dynamical
phenomena persist for different initial states and inhomogeneous Rabi-couplings.  A possible
experimental probe of our predictions and its feasibility are discussed in section VI.  In section
VII we summarize our results and provide an outlook.  Appendix A addresses our numerical methodology
based on MCTDHF. The numerical implementation of the single-shot simulations for spinor fermions is
briefly discussed in appendix B.  Finally, in appendix C we derive the effective spin-chain model
for our system.

\section{SETUP}
We consider an interacting system of $N$ spin-1/2 fermions of mass $m$, confined in an
one-dimensional parabolic trap of frequency $\omega$. The interparticle interaction emanating in
such systems is well-described by the $s$-wave contact interaction of strength, $g$ \cite{Olshanii}.
The latter can be manipulated by exploiting either Fano-Feshbach or confinement induced resonances
\cite{Chin}.  
The MB Hamiltonian that models such a system reads $\hat{\tilde H}=\hat H_\text{0}+\hat H_{I}$,
where the single-particle Hamiltonian $\hat H_0$ is
\begin{equation}
\hat H_{0}=\sum_{\alpha \in \{\uparrow,\downarrow\}} \int {\rm d}x~ \hat{\psi}^\dagger_\alpha(x) \left(-\frac{\hbar^2}{2 m} \frac{{\rm d}^2}{ {\rm d}x^2} +\frac{1}{2} m \omega^2 x^2 \right)\hat{\psi}_{\alpha}(x).\\
\end{equation}
The corresponding interparticle interaction term is encoded by
\begin{equation}
\hat H_{I}=g \int {\rm d}x~\hat{\psi}^\dagger_\downarrow(x) \hat{\psi}^\dagger_\uparrow(x) \hat{\psi}_\uparrow(x) \hat{\psi}_\downarrow(x),
\end{equation}
where $\hat{\psi}_{\alpha}(x)$ denotes the fermionic field-operator with spin $\alpha \in
\{\uparrow,\downarrow\}$.
The Hamiltonian $\hat{\tilde H}$ exhibits several crucial spin-symmetries. It can be shown that
 $\hat{\tilde H}$ commutes with each component of the total spin-vector operator
\begin{equation}
        \hat{\bm{S}} = \frac{\hbar}{2} \sum_{\alpha \alpha'} \int {\rm d}x ~
        \hat{\psi}^\dagger_\alpha(x) \bm{\sigma}_{\alpha,\alpha'} \hat{\psi}_{\alpha'}(x),
\end{equation}
where $\bm{\sigma}$ denotes the Pauli vector. Additionally, it possesses a SU($2$) symmetry stemming from its commutation with the total
spin-magnitude operator
\begin{equation}
\hat{S^2}= \frac{3 N\hbar^2}{4} +\underbrace{\frac{\hbar^2}{4}\sum_{\alpha, \alpha',
      \beta, \beta'} \int {\rm d}^2x~ \left( \bm{\sigma}_{\alpha,\alpha'} \cdot \bm{\sigma}_{\beta,\beta'} \right)
\hat{\psi}^\dagger_{\alpha}(x_1) \hat{\psi}^\dagger_{\beta}(x_2)  \hat{\psi}_{\beta'}(x_2)
\hat{\psi}_{\alpha'}(x_1)
}_{=\sum_{i \neq j} \hat{\bm{S}}_i \cdot \hat{\bm{S}}_j}.
\end{equation}
These symmetries imply that the eigenvalues of $\hat S_z$ and $\hat S^2$ define good quantum numbers.
Consequently, the ferromagnetic fully spin-polarized state, $|
\Psi_F \rangle= \prod_{i=0}^{N-1} \int {\rm d}x~ \phi_i(x) \hat \psi^\dagger_\uparrow (x) | 0 \rangle$,
where $\phi_i(x)$ refers to the $i$-th eigenfunction of the 1D harmonic oscillator, is the
energetically lowest eigenstate
of $\hat{\tilde H}$ (note that $\hat H_I | \Psi_F \rangle =0$) with total spin eigenvalues $S= S_z=\frac{N}{2}$.

To controllably probe the stability of such a ferromagnetic state $| \Psi(0) \rangle=| \Psi_F
\rangle$ under coherent processes that break both $\hat{S}_z$ and $\hat{S}^2$ symmetries we employ
an inhomogeneous Rabi coupling between the spin $\uparrow$ and $\downarrow$ states.  Note here that
similar Rabi-coupling techniques have been employed in several experiments e.g. see Ref.
\cite{ref3_1,ref3_2,ref3_3} involving binary bosonic mixtures.  The resulting Hamiltonian of the
total system reads 
\begin{equation}
    \hat H=\hat{H}_0+\hat H_S+\hat H_I,
\end{equation}
where the externally-imposed Rabi coupling term is
\begin{equation}
\hat H_{S}= % \frac{B_0}{\sqrt{2 \pi} w}  
B_0 \sum_{\alpha,\alpha'} \int dx~e^{-\frac{x^2}{2 w^2}} \hat{\psi}^\dagger_\alpha(x) \sigma^x_{\alpha,\alpha'} \hat{\psi}_{\alpha'}(x).
\end{equation}
In particular, $\hat H_S$ induces spin-flip transitions
with a spatially dependent transition amplitude, modelled by a Gaussian of width $w$ and
intensity $B_0$. This coupling scheme can be realized in ultracold experiments by
optical Raman dressing of the two lowest hyperfine levels of ${}^{40}$K (see also Section VI).
We choose the values $w=2$ and $B_0 = 2.5/\sqrt{8\pi}$ (in harmonic oscillator
units) leading to an average precession (Larmor) frequency, $\omega_L \approx 0.85$, for the spins which is lower than
all collective mode frequencies (e.g. $\omega_R \approx 2$ for the
breathing mode). This choice enables us
to avoid spin segregation phenomena \cite{AMR1} occurring when the length scale of the modulation $w$
is smaller than that of the trap $l_{\omega}=\sqrt{\hbar/m\omega}$.
%The perturbative nature of this coupling stems from the
%fact that the Larmor frequency for each of the spins deviates at most $12\%$ from its average value.

Our goal is to inspect the stability of the ferromagnetism when the $\hat H_S$ term, Eq. (6),
is abruptly switched on at $t=0$. To achieve this we track two main
observables, directly related with the system's broken symmetries. Namely, the
normalized spin polarization magnitude $P_S^{(1)}=\frac{2}{\hbar N}| \langle
\hat{\bm{S}} \rangle |$ and the spin-spin correlator $C^{(2)}_S = \frac{4
\langle \hat{S}^2 \rangle -3 N}{\hbar^2 N(N-1)}= \frac{\sum_{i \neq j} \langle
\hat{\bm{S}}_i \cdot \hat{\bm{S}}_j \rangle}{\hbar^2 N(N-1)} $.  $P^{(1)}_S$
expresses the averaged spin-order (magnetization) and refers to the magnitude
of the polarization. Due to its one-body character $P^{(1)}_S$ does not probe
the correlations that might emerge in the system despite being affected by them.
For this purpose we employ the spin-spin correlator, $C^{(2)}_S$, which probes
the alignment of each two spins and serves as an indicator for the distinction
of ferromagnetic $C^{(2)}_S \approx 1$, antiferromagnetic $C^{(2)}_S \approx
-1$ and paramagnetic $C^{(2)}_S \approx 0$ spin-spin correlations.

\section{ANALYSIS OF THE SPIN-FLIP DYNAMICS}
\begin{figure}[h]
    \includegraphics[width=0.8\columnwidth]{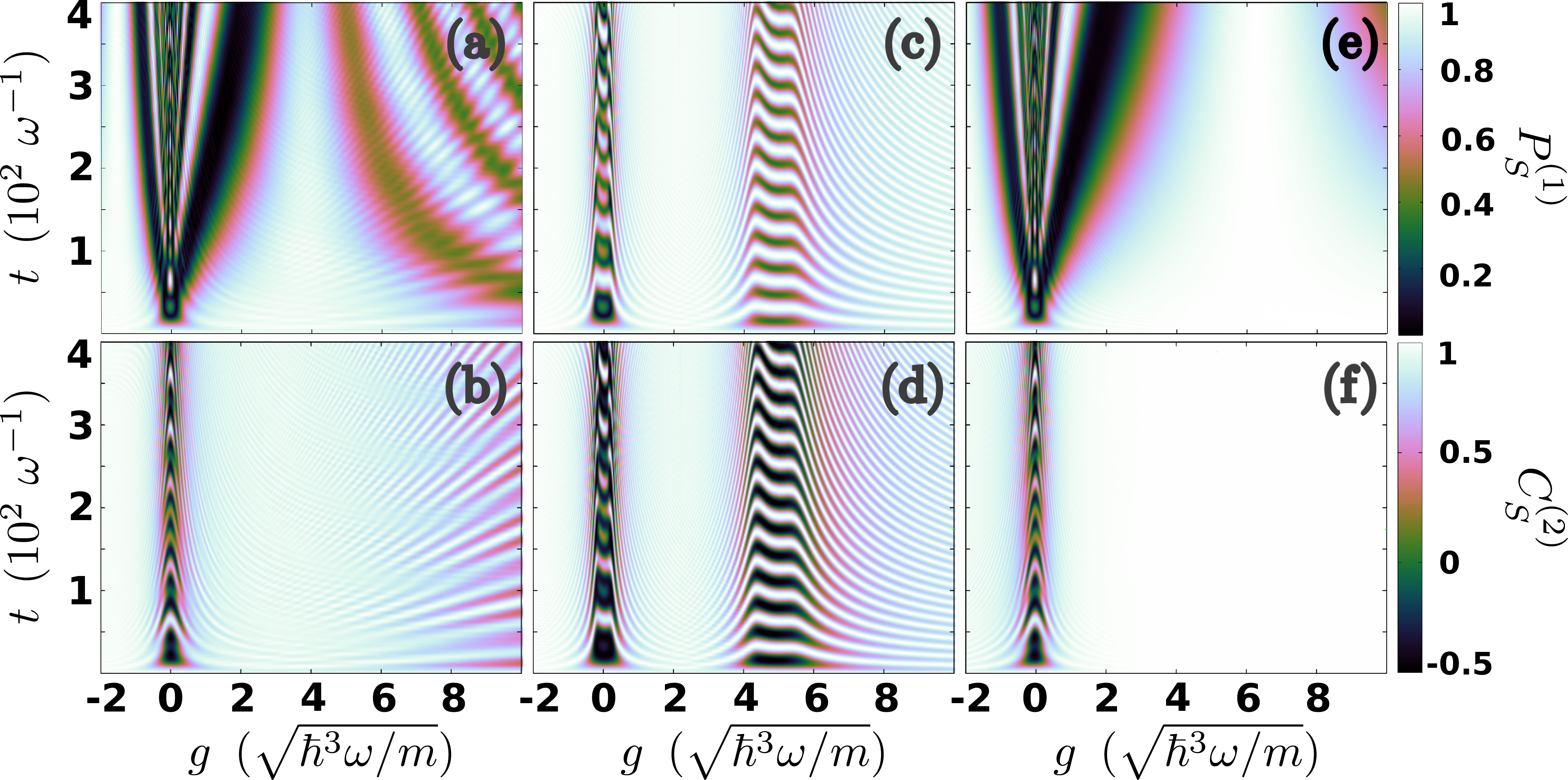}
    \caption{Time evolution of (a), (c), (e) the spin polarization magnitude $P_S^{(1)}$ and
        (b), (d), (f) the spin-spin correlator $C^{(2)}_S $ for varying $g$. The presented results refer to $N=3$ fermions by employing
        (a), (b) MCTDHF, (c), (d) TDHF and (e), (f) the generalized spin-chain approach.}
    \label{fig:mean_field}
\end{figure}

\subsection{Many-Body Correlated Spin-Flip Dynamics}
Figures 1(a) and 1(b) present our MB results for the paradigmatic case of $N=3$
fermions obtained via MCTDHF, that enables us to capture all interparticle correlation effects
\cite{Alon,Axel,MLX,Jenny1,Jenny2,Fpolarons}.
The MCTDHF method is a variational, numerically
exact, {\it ab initio} method
for solving the time-dependent MB Schr{\" o}dinger equation which includes all correlation effects. It is based on
expanding the MB wavefunction in terms of a time-dependent variationally
optimized basis. In this way it enables us to efficiently truncate the 
MB Hilbert space relevant for our system by using a computationally
feasible basis size. The MCTDHF method exhibits increased numerical efficiency
when compared to an expansion relying on a time-independent basis since the
number of basis states can be significantly reduced. A detailed discussion on the capabilities and the MB wavefunction ansatz of the above-mentioned method is presented in Appendix A.

For weak repulsive or attractive interactions, $|g|<
0.5$, a rapid demagnetization (see the decaying behaviour of
$P_S^{(1)}$) is observed,  accompanied by a loss of the spin alignment,
$C^{(2)}_S$, at a time scale $\sim 20$. Partial revivals of both $P_S^{(1)}$ and $C^{(2)}_S$ appear over regular time intervals for later
times.  Our results for this interaction interval are compliant with the
spin-dynamics analyzed in \cite{AMR1,AMR2} and we
shall refer to this regime as the weak-$g$ demagnetization regime. Indeed within this regime each of the particles precesses with a
different Larmor frequency leading to the loss of the polarization magnitude $P^{(1)}_S$ after a few precession cycles. For
intermediate interactions of either sign, i.e. $0.5 <|g|< 4$, the decay and
revival of $P_S^{(1)}$ also occurs but at a drastically
increased time-scale (which increases further for larger $|g|$) when compared to the
weak-$g$ demagnetization regime. In contrast, $C^{(2)}_S$ indicates
that the spins are close to be maximally aligned (e.g. $C^{(2)}_S \ge 0.85$ for
$2 < g<4$ and $C^{(2)}_S \ge 0.95$ for $g \sim -2$ in Fig. 1(b)) throughout the evolution,
signifying ferromagnetic spin-correlations. Therefore, ferromagnetism is unstable in
this interaction interval as the polarization ($P^{(1)}_S$) of the
ensemble features large fluctuations despite the ferromagnetic order
captured by $C^{(2)}_S$ which is almost perfect. Hence, we refer to this regime as
ferromagnetically ordered. In particular, it involves a different spin-order than ferromagnetism,
as its order is inferred by the ferromagnetic spin-spin correlations rather than the polarization. 
For $g > 4$  a suppression of the ferromagnetic
spin-spin correlations occurs as the amplitude of the $C^{(2)}_S$ oscillations increases for stronger $g$, see Fig. 1(b). For instance, at $g\approx 10$,
$C^{(2)}_S$ fluctuates between the values $0.5$ and unity.  $P^{(1)}_S$  is also
oscillatory taking values between unity and $1/3$, with a significantly smaller
oscillation frequency than $C^{(2)}_S$ [see Fig. 1(a)]. 
In the
following this interaction interval ($g >4$) is referred to as the strong-$g$ demagnetization
regime.

\begin{figure}[t]
    \includegraphics[width=0.8\columnwidth]{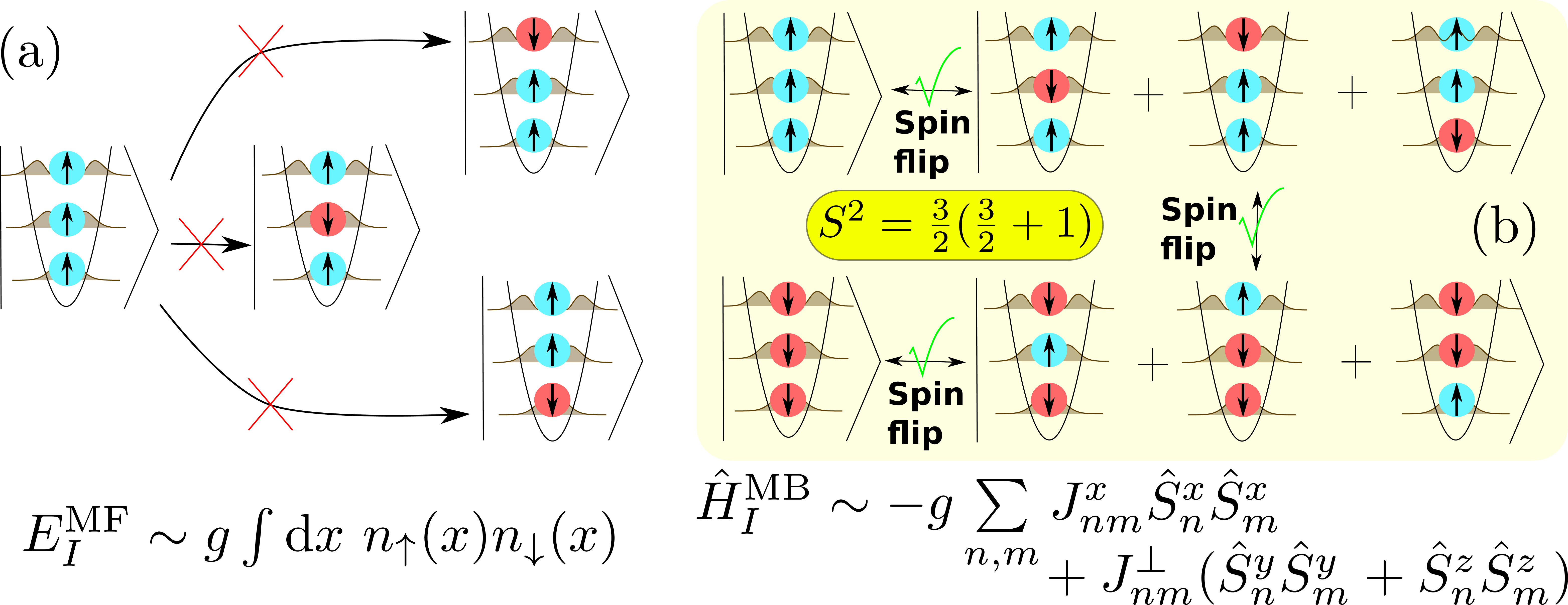}
    \caption{Spin-flip transitions of a spin-polarized few-fermion gas. (a) According to HF each $N$-body
        state corresponds to a single $N$-body (determinantal) configuration sketched here by the kets. For increasing
        interaction strength $g$, states with anti-oriented spins\footnote{For small interaction strength, $g$, and spin-flip coupling we can neglect paired states due to their large
         energy offset from the spin-polarized one.} accumulate large
        interaction energies $E_I^{\rm HF}$ that forbid spin-flip transitions from the fully polarized state
        (see crossed arrows). (b) In the MB case the interaction part of the Hamiltonian $H_I^{\rm
        MB}$ for small spin-flip coupling and $g$ approaches a XXZ model \cite{AMR1}, which
        allows for spin-flip transitions between all states with maximal total spin $S^2$.}
    \label{fig:sys}
\end{figure}

\subsection{Spin-Flip Dynamics Within TDHF} 
To demonstrate the crucial role of correlations
within the MB dynamics we compare the above MB findings with the TDHF
approximation presented in Fig.  1(c) and 1(d). For weak $|g|$ ($|g|<0.5$) the
demagnetization dynamics is qualitatively captured by the TDHF approximation.
However, upon increasing $|g|$, (in particular for $0.5<|g|<4$) TDHF predicts no loss of $P^{(1)}_S$
 in contrast to the MB case [compare Fig. 1(a) and 1(c)], while a
similar spin-correlation dynamics is observed [compare Fig. 1(b) and 1(d)].
This behaviour of the TDHF can be interpreted in terms of the Stoner model \cite{Stoner,Salasnich}, see Fig. 2(a).  Indeed, within HF the
    interaction energy of contact interacting spin-$1/2$ fermions is
    proportional to the density overlap between the two
spin-components \cite{Stoner,Salasnich}. Therefore, for large enough $g$ the system initialized in a
spin-polarized state characterized by zero interaction energy, cannot access states with a single (or more) spin-flips
due to their
large interaction energy. Thus, each of the spins has to preccess with the same
frequency resulting in the constant polarization magnitude, $P^{(1)}_S$. For strong $g>4$, Rabi oscillations between the
ferromagnetic initial state $| \Psi_F \rangle$ (characterized by $P^{(1)}_S=1$,
$C^{(2)}_S=1$) and the paired HF ground states $|\Psi^{\rm HF}_{\alpha} \rangle
\sim \int {\rm d}^3 x~ \phi_0(x_1) \phi_0(x_2) \phi_1(x_3)
\psi^\dagger_\uparrow(x_1) \psi^\dagger_\downarrow(x_2)
\psi^\dagger_\alpha(x_3) | 0 \rangle$ (referring to $P^{(1)}_S=1/3$,
$C^{(2)}_S=-1/2$) take place. This indicates that
the interparticle repulsion between the paired fermions exactly balances the
energy benefit of their pairing in the same state and corresponds to the Stoner
instability of the ground state. It is important to note here that these Rabi oscillations are absent in
the MB case, see also Fig. 1(c) and 1(d) for $g \approx 5$.

Concluding, the
ferromagnetically ordered regime exhibited in the MB case corresponds to a
stable ferromagnetic one within the HF framework. This observation exposes the correlated nature of
the ferromagnetically ordered regime. In both cases the interaction regime
is limited to intermediate values of $g$ and in particular to $0.5 < |g| <4$  [see Fig. 1(b), 1(d)].  However, for $g>4$ the mechanism
that breaks the spin-order differs. In the HF case the ground state Stoner
instability takes place which is forbidden for any finite repulsive interaction in
the MB case \cite{Lieb,Sowinski}. Instead, it is known that the unpaired states of maximum $S$ [see
Fig. 1(b)] and the MB antiferromagnetic
ground states exhibit a crossing in the Tonks-Girardeau limit 
\cite{ZurnAntiferro,Blume,BrouzosFer,Buignion,Lewenstein,Zinner}. In our case due to the breaking of
the SU(2) symmetry the above consist an avoided crossing
\cite{Cui_ref1,Ref1_1,Ref1_2} which is approached for increasing $g>4$. As we shall
argue in section IIID, the fluctuations of $C^{(2)}_S$ in the strong-$g$ demagnetization regime for the MB case can be
attributed to this avoided crossing.
\subsection{Effective Spin-Chain Model}
To uncover the main mechanisms responsible for the emergence of the ferromagnetically ordered regime ($0.5<|g|<4$)
we employ an extended version of the spin-chain model (see appendix C) presented in Ref.
\cite{AMR1}. Our spin-chain model 
incorporates additional effective magnetic field terms (for details see appendix C) when
compared to \cite{AMR1}, that are necessary for the treatment
of generic spin-$1/2$ fermion systems with a single conserved spin-component (here $\hat S_x$).
Within this model the $N$-body wavefuntion is decomposed as
    $|\Psi(t)\rangle=\sum_{\vec{n}} |\Psi_{\vec{n}}(t)\rangle$ with
$|\Psi_{\vec{n}}(t)\rangle=\sum_{\vec{\alpha}} A_{\vec{n};\vec{\alpha}}(t) \hat c_{n_1 \alpha_1}^{\dagger} \dots \hat c_{n_N \alpha_N}^{\dagger} |0\rangle$.
The operator $\hat c_{n \alpha}^{\dagger}$ creates a fermion in the eigenstate $\ket{\chi_{n}^\alpha}= \int dx~ \chi^\alpha_{n}(x) \hat
\psi^\dagger_\alpha(x) \ket{0}$ of $\hat H_0+\hat H_S$, see Eq. (1) and (6). $\vec{n}=(n_1,\dots,n_N)$ and
$\vec{\alpha}=(\alpha_1,\dots,\alpha_N)$ denote the
spatial and spin configuration respectively.
The crucial approximation in this model is that all the interaction terms, $\hat H^{\delta \vec{n}}_I$, that couple different
spatial configurations $\vec{n}$ can be neglected. 
By setting $\hat H_I^{\rm eff} \equiv
\hat H_I -\hat H^{\delta \vec{n}}_I$, the different spatial configurations $\vec{n}$ are decoupled.
It can be shown that the time-evolution of each $|\Psi_{\vec{n}}(t)\rangle$
can be described by an XXZ spin-chain consisting of $N$-spins \cite{AMR1}. The employed approximation limits the expected range of validity of the spin-chain
model to small interaction values, $g < \sqrt{\hbar \omega^3/m}=1$, where the interaction energy
is smaller than the energy spacing between the single-particle eigenstates.

The polarization dynamics within the spin-chain model comply with the MB results within the weak-$g$
demagnetization ($|g|<0.5$) and the ferromagnetically ordered regime ($0.5<|g|<4$) [compare Fig.
1(e) and Fig. 1(a)].  Moreover, by comparing $C^{(2)}_S$ [see Fig. 1(b) and 1(f)] between the two
methods we observe that the spin-correlation dynamics is almost identical in the weak-$g$
demagnetization regime, where the approximations that the spin-chain model employs are valid. On the
contrary, in the ferromagnetically ordered regime the ferromagnetic spin-correlations are
overestimated by the spin-chain method [hardly visible in Fig. 1(b) and 1(f)]. Finally, for
increasing interactions ($g>4$) no strong-$g$ demagnetization regime appears within the spin-chain
model, signifying the break down of its validity. This behaviour is clearly imprinted in $C^{(2)}_S$
for large $g$ [compare Fig. 1(f) and 1(b) for $g \approx 8$].  To interpret the spin-chain dynamics
in the ferromagnetic and strong-$g$ demagnetization regime we note that the configuration
$\vec{n}=(0,1,2)$ possesses approximately $99.72 \%$ of the contribution to $| \Psi(0) \rangle$ and
thus it almost completely dictates the dynamics of the system within the spin-chain approximation.
The MB polarization $P^{(1)}_S$ dynamics within the ferromagnetically ordered regime is
well-captured by the spin-chain model allowing us to conclude that this behaviour emerges due to the
spin-dynamics of the different states (characterized by distinct $\vec{\alpha}$) within the dominant
$\vec{n}=(0,1,2)$ spatial configuration. In contrast, the (small) depletion of $C_S^{(2)}$ in the
same regime [see Fig. 1(b)] is absent in the spin-chain approximation, leading to the conclusion
that it stems from the neglected couplings to different spatial configurations, contained in $\hat
H_I^{\delta \vec{n}}$.  The latter couplings, however, are not as strong as to prohibit the
spin-chain model to capture the spin-order emerging within the MB evolution in this interaction
regime.  Regarding the absence of a strong-$g$ demagnetization regime we remark here that the
coupling between the antiferromagnetic ground states belonging to the spatial configuration
$\vec{n}=(0,0,1)$ and the initially populated states of the dominant $\vec{n}=(0,1,2)$ configuration
is neglected by the spin-chain model.

At this point it becomes clear that the ferromagnetic order exhibited in 1D spin-1/2 Fermi gases
greatly deviates from the
standard HF description. Additionally, the emerging spin-order is different than 
the one perceived as ferromagnetic in the literature. Indeed, its defining characteristic is the stability of the spin-spin correlations rather than the
polarization. The spin-chain model seems to capture well some of the characteristics of this emerging order.
In the following by analyzing in parallel the MB and spin-chain dynamics we will shed light onto the
underlying microscopic mechanisms of the ferromagnetically ordered regime.

\subsection{Analysis Of The Microscopic Mechanisms}

\begin{figure}[h]
    \includegraphics[width=0.9\columnwidth]{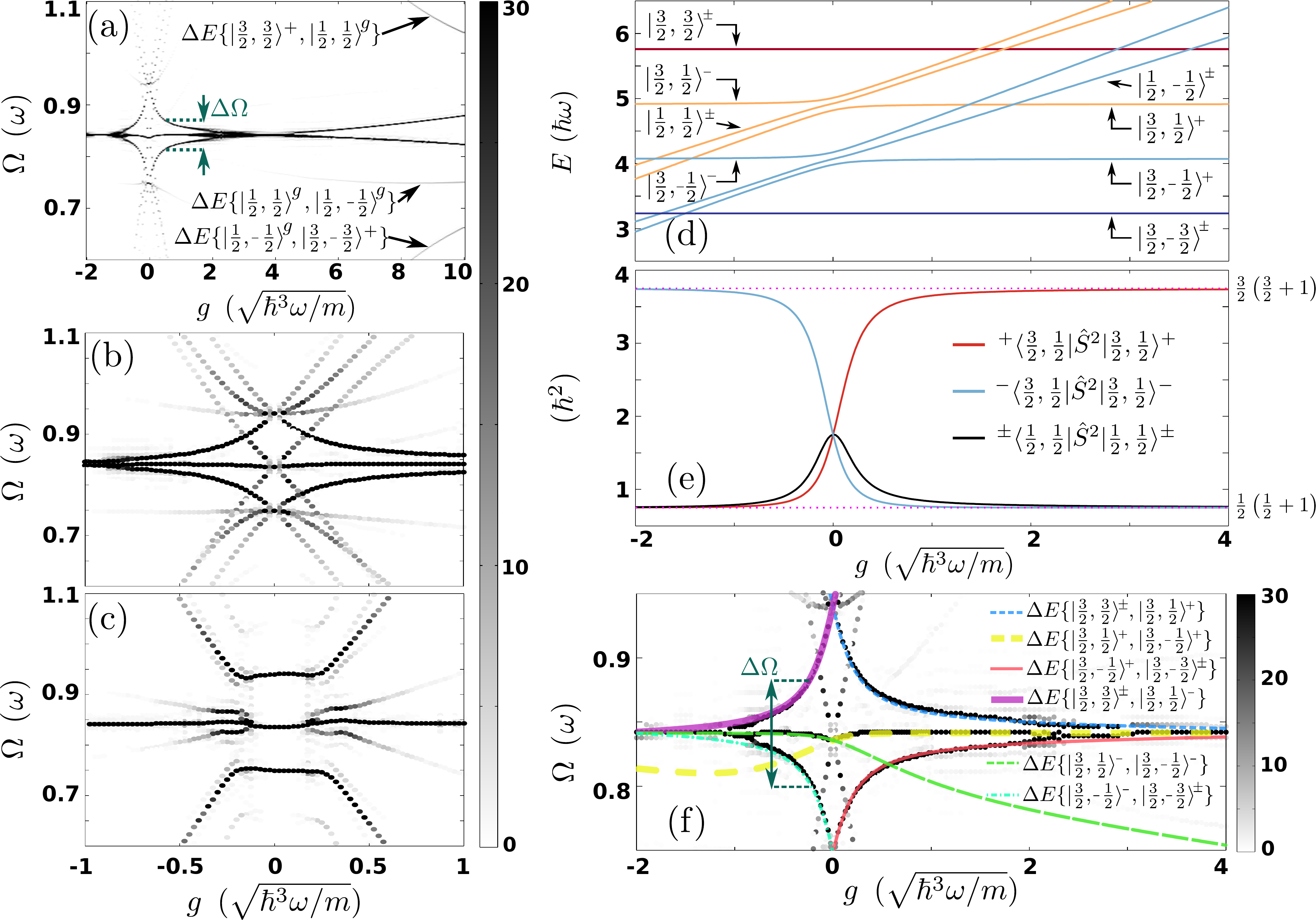}
    \caption{(a), (b) Spectrum of the polarization vector $\mathcal{F} \{P_S^{(1)} \}$ for $N=3$ fermions within MCTDHF.
    $|\frac{1}{2},\pm \frac{1}{2}\rangle^g$ refers to the antiferromagnetic ground states.
    (c) $\mathcal{F} \{P_S^{(1)} \}$ within TDHF. (d) Eigenspectrum for all states within
    the $\vec{n}=(0,1,2)$ configuration within the spin-chain model. (e) Expectation value of
    the $\hat{S}^2$ operator for selected
    eigenstates of the spin-chain model. The dotted lines indicate the value of $\langle \hat{S}^2
\rangle$ for $S=\frac{1}{2}$ and $S=\frac{3}{2}$.
         (f) Comparison of (a) with the eigenenergy differences within the spin-chain model.}
    \label{fig:mean_field}
\end{figure}

To identify the underlying mechanisms of the MB spin-dynamics we invoke the spectrum of the spin-polarization, namely
$\mathcal{F} \{ P_S^{(1)} \}(\Omega)=\big|\int
{\rm d}t ~e^{i \Omega t} P^{(1)}_S (t)\big|^2$, presented in Fig. 3(a) and 3(b) for $-2<g<10$ and $|g|<1$ respectively.
Recall that each branch in the spectrum of $P_S^{(1)}$ corresponds to an energy difference between
two eigenenergies of $\hat H$, [see Eq. (5)].
For $g=0$ three distinct Larmor frequencies\footnote{Note here the perturbative nature of $\hat
    H_S$, Eq. (6), which is manifested by the fact that the Larmor frequency of the occupied states does
not deviate more than $12\%$ from its average value $\Omega_L\approx 0.85$.} occur
that correspond to the three energy differences among the occupied single-particle eigenstates in
the spatial configuration $\vec{n}=(0,1,2)$.
For $g \neq 0$ a
multitude of interaction-dependent frequency branches emerge from each Larmor frequency. 
The failure of TDHF to capture even on the qualitative level the spin dynamics even for low $g$ is evident in $\mathcal{F} \{ P_S^{(1)} \}(\Omega)$.
Indeed, the TDHF captures only one frequency per particle for $|g|<0.1$ [see Fig. 3(c)] in contrast to the 
multitude of interaction-dependent frequency branches emerging from each Larmor frequency in the correlated case [see Fig. 3(b)].
Thus we can conclude that the build up of correlations in the MB case even for very small $g$ completely invalidates the HF picture for
the spin dynamics. Such correlations are of particular importance in the ferromagnetically ordered regime. In this case, three dominant branches appear in the vicinity of $\Omega \approx 0.85$
[see also Fig. 3(b) for $g \approx 1$] within the correlated case that lead to the beating dynamics of $P^{(1)}_S$, observed in Fig. 1(a), 1(b).
% The remaining
% frequencies appearing in this regime [Fig. 3(b)] are negligible as they correspond to eigenenergy differences between states
% of different $S$, which become energetically well-separated due to the interaction.

The origin of the above-mentioned frequencies can be exposed by comparing 
$\mathcal{F}\{ P_S^{(1)} \}$, with the energy differences of the eigenstates of
the spin-chain model. As anticipated by our discussion in section IIIC the
eigenstates of the configuration $\vec{n}=(0,1,2)$ are expected to well capture the
$P^{(1)}_S$ dynamics. The eigenspectrum of the spin-chain Hamiltonian $\hat
H_{\vec{n}}=\hat H_{(0,1,2)}$ is presented in Fig.  3(d). The spin-chain
eigenstates within the weak-$g$ demagnetization regime ($|g| <0.5$) are ordered in terms of
increasing $S_x$ due to the Zeeman effect induced by the effective magnetic
field along the $x$ axis that $\hat H_S$, Eq. (6), introduces. Within this weak-$g$ demagnetization
regime an avoided crossing between the distinct eigenstates of the same
$S_x=\pm \frac{1}{2}$ occurs which can be attributed to the breaking of the $\hat
S^2$ symmetry by the inhomogeneity of the $\hat H_S$ term.  The states of the
spin-chain model can be labeled as $|S,S_x\rangle^\ell$ in the repulsive ($\ell
= +$) and attractive ($\ell = -$) part of the ferromagnetically ordered regime ($0.5<|g|<4$).
This is possible because $S$ is an approximate good quantum number within this
interaction range. This property can be identified by examining the expectation
value of $\langle \hat{S}^2 \rangle\equiv {}^{\ell}\langle S,S_x|\hat{S}^2| S,S_x \rangle^{\ell}$.  Figure 3(e) presents this
expectation value for the states with $S_x=\frac{1}{2}$ and varying interaction
strength.  It becomes evident that for increasing $|g|$ the $\langle \hat S^2
\rangle$ of one of these states [i.e. $|\frac{3}{2},\frac{1}{2}\rangle^{+}$
and $|\frac{3}{2},\frac{1}{2}\rangle^{-}$ in the case of $g>0$ and $g<0$
respectively] approaches the value $\langle \hat
S^2 \rangle=\frac{3}{2}\langle \frac{3}{2}+1 \rangle$, indicating that $S\approx \frac{3}{2}$. Moreover, the states
$\{|\frac{3}{2},\frac{1}{2}\rangle^{-}, |\frac{3}{2},\frac{1}{2}\rangle^{\pm}\}$ and
$\{|\frac{3}{2},\frac{1}{2}\rangle^{+}, |\frac{3}{2},\frac{1}{2}\rangle^{\pm}\}$ in the case of $g>0$ and $g<0$
respectively tend to $\langle \hat S^2 \rangle=\frac{1}{2}( \frac{1}{2}+1 )$. We remark here that the states with $S_x=-\frac{1}{2}$ exhibit a similar
behavior as the aforementioned $S_x=\frac{1}{2}$ case and the eigenstates $|\frac{3}{2},\pm \frac{3}{2} \rangle^{\pm}$ correspond to fully
polarized states along the $x$ axis with $\langle \hat
S^2 \rangle=\frac{3}{2}\langle \frac{3}{2}+1 \rangle$ for all $g$ (not shown here for brevity).
By employing the spin-chain eigenspectrum, Fig. 3(d), we can identify all the energy branches appearing in Fig. 3(b) with the
corresponding eigenenergy differences of the spin-chain model. Most importantly, Fig. 3(f)
presents this identification within the ferromagnetically ordered regime ($0.5<|g|<4$).
As it can be seen the energy differences between the spin-chain
eigenstates with $S\approx \frac{3}{2}$, $\Delta E \{|\frac{3}{2},S_x
\rangle^{\ell}, |\frac{3}{2},S'_{x} \rangle^{\ell'}\}$, possess the main
contribution to $\mathcal{F} \{ P_S^{(1)} \}$ in the ferromagnetically ordered
regime as they match well with the dominant energy branches appearing in the MB dynamics. Based on
this identification we can draw several conclusions regarding the order exhibited
within the ferromagnetically ordered regime. First, the ferromagnetic
spin-correlations emanate from the predominant occupation of MB eigenstates
with $S \approx 3/2$ each characterized by a different value of $S_x$. The frequency difference
$\Delta \Omega$ between the highest ($\Delta E
    \{|\frac{3}{2},\frac{3}{2}\rangle^\pm,|\frac{3}{2},\frac{1}{2}\rangle^l$)
    and lowest lying ($\Delta E
        \{|\frac{3}{2},-\frac{1}{2}\rangle^l,|\frac{3}{2},-\frac{3}{2}\rangle^\pm$)
        of the above-mentioned dominant branches [see also Fig. 3(d)] results
        in the dephasing of the initial superposition. In terms of the spin-$\uparrow$ and spin-$\downarrow$ states this
        means that the system performs spin-flips between the different states with
    $S=\frac{3}{2}(\frac{3}{2}+1)$ and varying $S_z$ leading to the decay of $P^{(1)}_S$, see Fig.
2(b). The corresponding timescale is $\tau_D
        \approx \frac{\pi}{\Delta \Omega}$. $\Delta \Omega$ is attributed to
        the energy shift of the $| \frac{3}{2},\pm \frac{1}{2}\rangle^\ell$
        states from being equidistantly spaced which is induced by an avoided
        crossing at $g=0$, as in Fig. 3(d).  The increase of $|g|$
        within the ferromagnetically ordered regime, leads to a decrease of
        $\Delta \Omega$ (or equivalently increase of $\tau_D$) giving rise to
        the observed $P_S^{(1)}$ dynamics [Fig. 1(a) and 1(e)]. 

Regarding the strong-$g$ demagnetization regime [see Fig. 3(a) for $g>4$] the frequency branches that correspond to the 
$|\frac{3}{2},\pm \frac{1}{2}\rangle^+$ states deviate from $\Omega \approx 0.85$ as 
they couple to the
antiferromagnetic ground states of $S \approx \frac{1}{2}$ and $S_x=\pm \frac{1}{2}$ character. Due to the
same mechanism additional branches also
appear in $\mathcal{F} \{P^{(1)}_S\}$ [see the corresponding arrows in Fig. 3(a)] that contribute to the
complex $P^{(1)}_S$ dynamics exhibited in this interaction regime [see Fig. 1(a)] and results in the
oscillatory patterns of $C_S^{(2)}$ [see Fig. 1(b), and also our discussion in section IIIB].

\begin{figure}[t]
    \includegraphics[width=0.8\columnwidth]{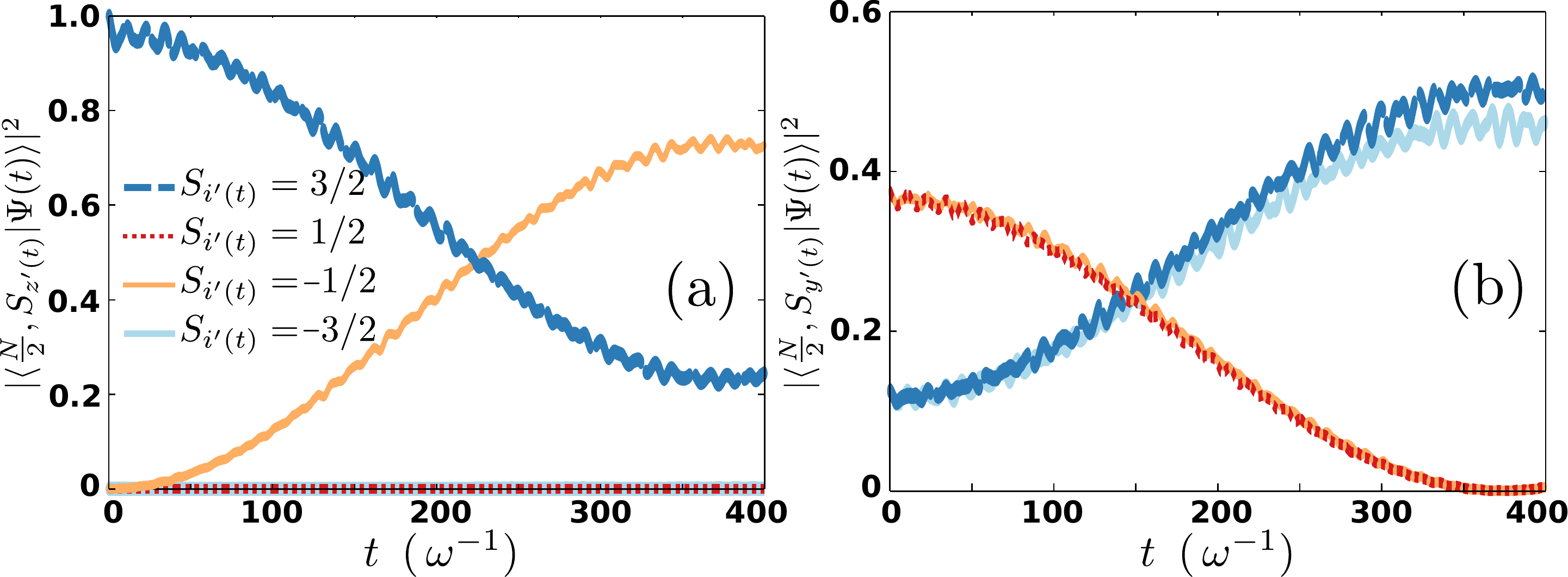}
    \caption{Time evolution of the populations of the $|\frac{3}{2},S_{i'(t)}\rangle$ states of
        the $\vec{n}=(0,1,2)$ configuration for $N=3$, $g=2$ within
    MCTDHF. (a) $i'(t)=z'(t)$ and (b) $i'(t)=y'(t)$.}
    \label{fig:ns_field}
\end{figure}

\subsection{Characterization of Entanglement In The Ferromagnetically Ordered Regime}
The separation of the energy scales $\Delta
\Omega \sim 0.05$ [see Fig. 3(c)] and the average Larmor frequency, $\Omega_L=\frac{1}{3 \hbar}\Delta E\{| \frac{3}{2},
\frac{3}{2} \rangle,| \frac{3}{2}, -\frac{3}{2} \rangle\} \sim 0.85$ observed in Fig. 3, enables us to further characterize the
superpositions that emerge in the ferromagnetically ordered regime. To this end we introduce the precessing bases $|\frac{3}{2},S_{j'(t)}\rangle=e^{i \frac{1}{2} \hat{S}_x\Omega_L t}|\frac{3}{2},S_{j}\rangle$ ($j
\in \{x,y,z\}$) and project the MB wavefunction obtained within MCTDHF to these basis states. Note that if all the particles were collectively precessing with the frequency $\Omega_L$ then $|\langle
\frac{3}{2},S_{z'(t)}=\frac{3}{2} |\Psi(t) \rangle|$ would be constant in time. 
However, as we have already enstablished in section IIIB, this is not the case. Fig. 4(a) presents the results of this projection for a representative case ($g=2$) within the ferromagnetically
ordered regime. We observe a low-frequency ($\sim \Delta \Omega$) population transfer
from the state $|\frac{3}{2},S_{z'(t)}=\frac{3}{2}\rangle$ to the state
$|\frac{3}{2},S_{z'(t)}=-\frac{1}{2}\rangle$.
For $t > 350$ the latter mechanism results in $\langle \hat{S}_{z'(t)} \rangle \approx 0$, as $|\frac{3}{2},S_{z'(t)}=-\frac{1}{2}\rangle$
possesses approximatively a three times larger population as compared to $|\frac{3}{2},S_{z'(t)}=\frac{3}{2}\rangle$. The
nature of this superposition can be understood by transforming to the orthogonal precessing axis, $y'(t)$ [see
Fig. 4(b)]. In this case the populations of
$|\frac{3}{2},S_{y'(t)}=\frac{3}{2}\rangle$ and
$|\frac{3}{2},S_{y'(t)}=-\frac{3}{2}\rangle$ are almost equal for $t>350$, signifying the
tendency to dynamically approach a NOON state, characterized by $\langle \hat{S}_{y'(t)} \rangle \approx 0$,
i.e. $|\Psi (t_0) \rangle \sim \frac{1}{\sqrt{2}} (
|\frac{3}{2},S_{y'(t)}=\frac{3}{2}\rangle + e^{i \phi} |\frac{3}{2},S_{y'(t)}=-\frac{3}{2}\rangle )$
with a relative phase $\phi$.
These results combined 
with the conserved quantity $\langle \hat S_x \rangle = 0$ explain the decay of the total
magnetization $P^{(1)}_S \to 0$ for $t > 350$. Accordingly,
the spin-dynamics within the ferromagnetically ordered regime describes the dynamical evolution of a
fully-polarized state to a superposition one consisting of two antiparallel-oriented fully-polarized states.

\section{SPIN DYNAMICS FOR $\mathbf{N=5}$ FERMIONS}

We next demonstrate the robustness of our main findings for the case of larger
particle numbers by examining a system consisting of $N=5$ fermions.
% The system parameters, namely $B_0=\frac{2.5}{\sqrt{8 \pi}}$ and $w=2$, are the same as in the
% $N=3$ fermion case (see main text).

\begin{figure}[ht]
    \includegraphics[width=1.0\columnwidth]{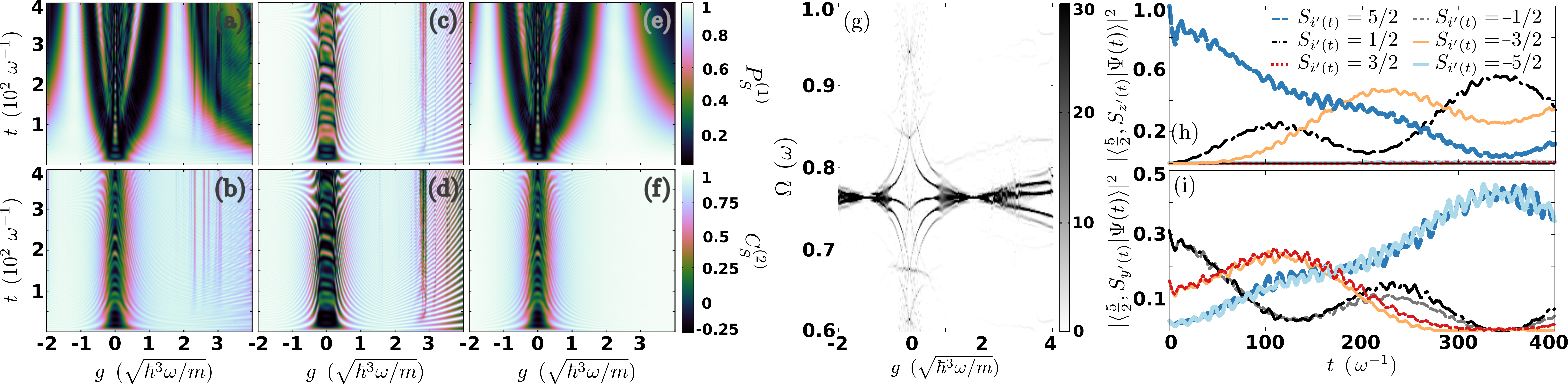}
    \caption{Time evolution of (a), (c), (e) the spin polarization magnitude $P_S^{(1)}$ and
        (b), (d), (f) the spin-spin correlator $C^{(2)}_S $ for varying $g$. The presented results refer to $N=5$ fermions by employing
        (a), (b) the MCTDHF, (c), (d) the TDHF and (e), (f) the generalized spin-chain approach. 
    (g) Spectrum of the polarization vector $\mathcal{F} \{P_S^{(1)} \}$ for $N=5$ fermions within
    MCTDHF.  Time evolution of the populations
    of the $|\frac{5}{2},S_{i'(t)}\rangle$ states of
the $\vec{n}=(0,1,2,3,4)$ configuration for $N=5$, $g=1$ within MCTDHF. (h) $i'(t)=z'(t)$ and (i) $i'(t)=y'(t)$.}
    \label{fig:sys_sup}
\end{figure}

Figures 5(a)-5(f) present
$P^{(1)}_S$ and $C^{(2)}_S$ for $N=5$ fermions within the tree different approaches employed above, i.e. the MCTDHF, TDHF and spin-chain approach.
A similar spin-dynamics as for the $N=3$ case is
observed for both quantities, but in different interaction regimes, caused by the increase of $N$.
The ferromagnetically ordered regime occurs in the range $0.5<|g|<2$, where both $P^{(1)}_S$ and
$C^{(2)}_S$ possess a value proximal to unity within the TDHF approach [see Fig. 5(c) and 5(d)]. In the
same range the MB treatment provided by MCTDHF reveals that $P^{(1)}_S$ is decaying [see
Fig. 5(a)], a feature which is also well captured by the spin-chain method [see Fig. 5(e)]. The only
additional structures that emerge in the MB spin-flip dynamics when compared to the $N=3$ case are
very narrow interaction windows where $C^{(2)}_S$ gets significantly depleted from unity [see Fig. 5(b),
$g \approx 2.5$]. These regions can be attributed to avoided crossings between the different spin-states of the
dominant spatial configuration $\vec n=(0,1,2,3,4)$ with states
characterized by spatial configurations with double occupations of single-particle spatial mode(s) [e.g. $\vec{n}'=(0,1,2,2,3)$].

Inspecting $\mathcal{F} \{ P^{(1)}_S \}$ for $N=5$ fermions, see Fig. 5(g), similar microscopic
mechanisms to the $N=3$ case can be observed in both the weak-$g$ and the ferromagnetically ordered
regime.  Despite the fact that more states are involved, the main features essentially remain the
same.  The weak-$g$ demagnetization regime originates from the multitude of branches emerging from
the five available non-interacting Larmor frequencies. However, only five (which can be identified
as the energy differences between the $|\frac{5}{2},S^x\rangle^\ell$ states) possess a significant
amplitude for $|g|>0.5$. The frequency difference, $\Delta \Omega$, between the highest and lowest
lying of the above five branches [see Fig. 5(g)] gives rise to the decay of $P_S^{(1)}$
within the ferromagnetically ordered regime observed in Fig. 5(a). 

Finally, we show that even the superpositions emerging in the dynamics are of the same character as for
$N=3$ fermions. To reveal this we construct the precessing basis analogously to the case $N=3$,
namely $|
\frac{5}{2}, S_j'(t) \rangle = e^{i \frac{1}{2} \hat{S}_x \Omega_L t} | \frac{5}{2}, S_j \rangle$,
$\Omega_L=\frac{1}{5\hbar} \Delta E\{| \frac{5}{2}, \frac{5}{2} \rangle, | \frac{5}{2},
-\frac{5}{2} \rangle \}$ and expand the MB wavefunction in terms of the latter. Figures 5(h) and
5(i) present the results of this expansion for the axes $z'(t)$ and $y'(t)$ respectively and for
$g=1$ within the ferromagnetically ordered regime. Figure 5(h) demonstrates that the collective precession of
the spins characterized by $| \langle \frac{5}{2}, S_{z'(t)}=\frac{5}{2}|\Psi(t)\rangle|^2=1$ gets
quickly dephased. At later times $t \approx 350$ the dephasing of the collective Larmor precession
leads to the formation of a NOON-like state characterized by $| \langle \frac{5}{2},
S_{y'(t)}=\frac{5}{2}|\Psi(t)\rangle|^2=| \langle \frac{5}{2},
S_{y'(t)}=-\frac{5}{2}|\Psi(t)\rangle|^2 \approx 1/2$ [see Fig. 5(i)], compliant with our $N=3$
results.

\section{GENERALIZATION TO OTHER DYNAMICAL SYSTEMS}
\begin{figure}[t]
    \centering
    \includegraphics[width=0.8\textwidth]{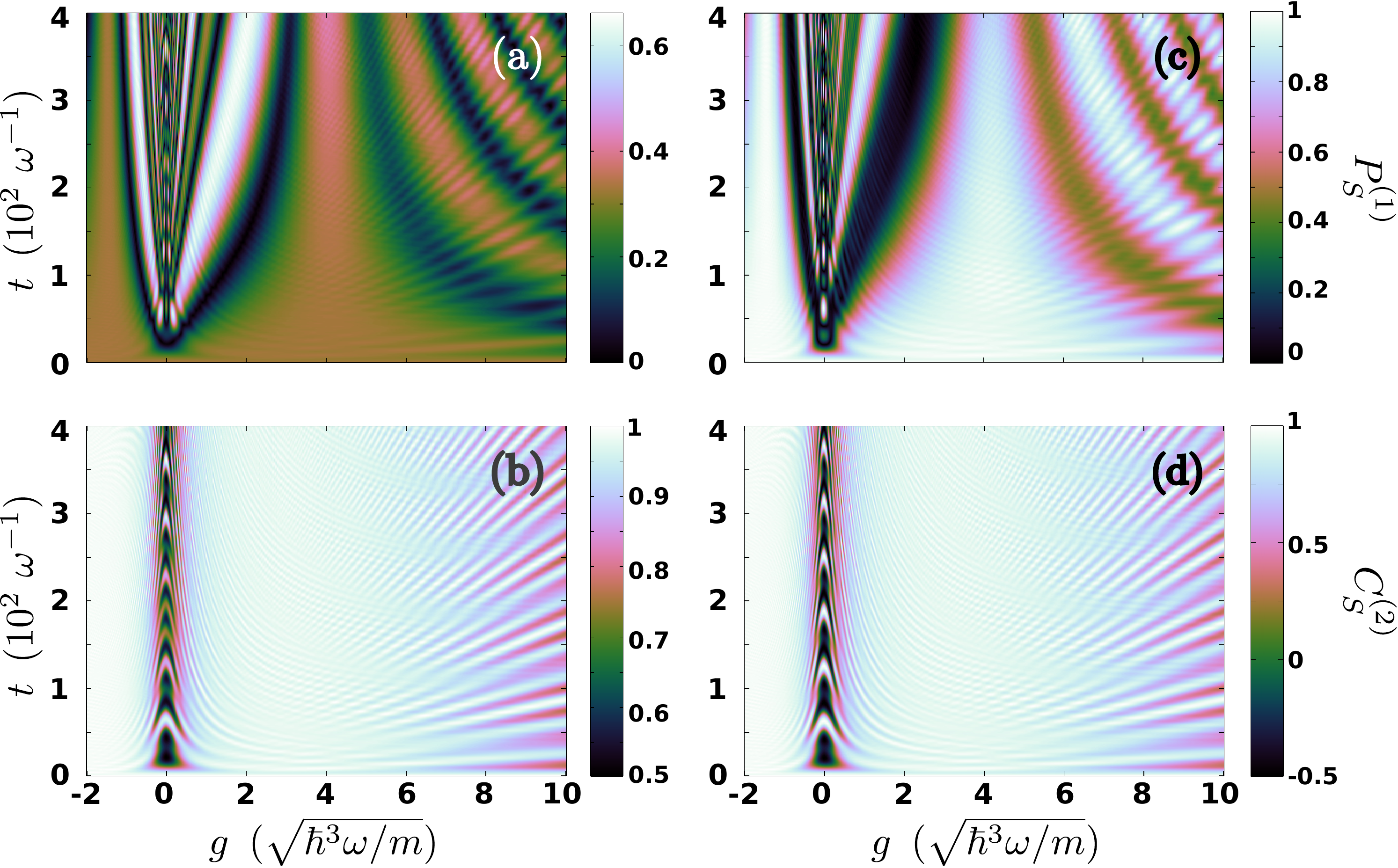}
    \caption{Time evolution of (a), (c) the spin polarization magnitude
        $P^{(1)}_S$ and (b), (d) the spin-spin correlator
        $C^{(2)}_S$ for varying $g$. The presented $N=3$ results are derived within MCTDHF
        (a), (b) with $|\Psi(0)\rangle=|\Psi_{S=3/2,Sz=1/2} \rangle$ as initial state (see text) and by
        employing the Hamiltonian $\hat H$ or (c), (d) with $|\Psi(0)\rangle=|\Psi_0 \rangle$ and
    $\hat{\tilde{H}} +\hat H'_S$.} \label{fig:sys2}
\end{figure}

Below we demonstrate that the above identified ferromagnetic properties are not restricted to the
previously examined  out-of-equilibrium scenario. Indeed, we will show that the 1D spin-$1/2$ Fermi
gas exhibits a similar spin-dynamics for different initial states characterized by $S\approx
\frac{N}{2}$ within the ferromagnetic-like regime, which, furthermore, does not depend on the exact
form of the Rabi coupling potential\footnote{Here we assume a weak and spatially slowly varying
coupling of the spin-$\uparrow$ and spin-$\downarrow$ states.}. The special feature of the specific
dynamical protocol investigated in the previous sections is that it can be readily implemented in
state-of-the-art experiments (see also section VI).

It can be shown that any initial state which is dominated by the states of the
$\vec{n}=(0,\dots,N-1)$ spatial configuration with $S=\frac{N}{2}$ possesses a similar spin-spin
correlation dynamics to the ferromagnetic one, $| \Psi_F \rangle=\prod_{i=0}^2 \int {\rm
d}x~\phi_i(x) \hat{\psi}^\dagger_\uparrow (x) |0 \rangle$. The reason is that the time-evolution of
the expectation value of $\hat S^2$ reads
\begin{equation}
    \langle  \Psi (t)| \hat{S}^2| \Psi (t) \rangle = \sum_{S_x=-\frac{N}{2}}^{\frac{N}{2}}
    \bigg|\Big\langle \Psi (t)\Big| \frac{3}{2}; S_x \Big\rangle\bigg|^2 \bigg\langle  \frac{3}{2}; S_x \bigg| e^{\frac{it}{\hbar}
    \hat{H}} \hat{S}^2 e^{-\frac{it}{\hbar} \hat{H}} \bigg| \frac{3}{2}; S_x \bigg\rangle
    +\mathcal{O}\left( 1- \sum_{S_x=-\frac{N}{2}}^{\frac{N}{2}}
    \bigg|\Big\langle \Psi (t)\Big| \frac{3}{2}; S_x \Big\rangle\bigg|^2\right).
    \label{expansion}
\end{equation}
This equation stems from the fact that the quantum numbers $S=\frac{N}{2}$ and $S_x \in
\{-\frac{N}{2},\dots,\frac{N}{2}\}$ uniquely identify a single $N$-body state of the
$\vec{n}=(0,\dots,N-1)$ configuration. Additionally, the $\hat{S}_x$ operator commutes with both
$\hat S^2$ and $\hat{H}$ canceling all cross terms that would appear in the first term of Eq.
(\ref{expansion}). Therefore, within the ferromagnetically ordered regime where $\langle  \Psi (t)|
\hat{S}^2| \Psi (t) \rangle \approx \frac{N}{2}\left( \frac{N}{2} + 1 \right)$, for $| \Psi(0)
\rangle=| \Psi_F \rangle$, all of the contributing expectation values need to satisfy $\langle
\frac{N}{2}; S_x | e^{\frac{it}{\hbar} \hat{H}} \hat{S}^2 e^{-\frac{it}{\hbar} \hat{H}} |
\frac{N}{2}; S_x \rangle \approx \frac{N}{2}\left( \frac{N}{2} + 1 \right)$, since the probabilities
are positive definite, i.e. $|\langle \Psi (t)| \frac{3}{2}; S_x \rangle|^2 >0 $. This implies that
for all initial states which satisfy $\sum_{S_x=-\frac{N}{2}}^{\frac{N}{2}} |\langle \Psi (0)|
\frac{3}{2}; S_x \rangle|^2 \approx 1$ the correlation dynamics within the ferromagnetically ordered
regime is stable.

To provide concrete numerical evidence supporting the above-mentioned theoretical argument, we
present in Fig. 6, the time-evolution of the polarization, $P^{(1)}_S$, and the spin-spin
correlator, $C^{(2)}_S$, utilizing the Hamiltonian of Eq. (5), in the case of the initial state
\cite{ZurnAntiferro}
\begin{equation}
    |\Psi_{S=3/2,Sz=1/2} \rangle=\int \frac{{\rm d}^3x}{\sqrt{3!}}
        \begin{vmatrix}
\phi_{0}(x_1) && \phi_{0}(x_2) && \phi_{0}(x_3)\\
\phi_{1}(x_1) && \phi_{1}(x_2) && \phi_{1}(x_3)\\
\phi_{2}(x_1) && \phi_{2}(x_2) && \phi_{2}(x_3)
    \end{vmatrix} 
 \hat \psi^{\dagger}_{\uparrow}(x_1)\hat \psi^{\dagger}_{\uparrow}(x_2)\hat
 \psi_{\downarrow}^{\dagger}(x_3) | 0 \rangle.
\end{equation}
As it can be clearly seen the correlation dynamics [see Fig 6(b)] is almost identical to the one
observed in section IIIA for the fully spin-polarized initial state, $| \Psi_F \rangle=\prod_{i=0}^2
\int {\rm d}x~\phi_i(x) \hat{\psi}^\dagger_\uparrow (x) |0 \rangle$ [see also Fig. 1(b)]. Most
importantly, the ferromagnetically ordered regime appears in the $0.5<|g|<4$ interaction range,
characterized by $C^{2}_S\approx 1$, while losses of spin-alignment (i.e.  $C^{2}_S<1$) are found
outside of this interaction regime. The polarization dynamics [see Fig 6(a)], however, shows
different patterns from the dynamics obtained with $|\Psi_F\rangle$, since the initial polarization
in the present case is $P^{(2)}_S=\frac{1}{3}$ rather than unity.  Nevertheless, within the
ferromagnetically ordered regime we observe large fluctuations of the polarization while the
spin-spin correlator is almost constant, similarly to the dynamics that the system with $| \Psi(0)
\langle = | \Psi_F \rangle$ follows [see Fig. 1 (a), (b)].

    According to our previous discussion (see Section II) the Rabi-coupling between the
    spin-$\uparrow$ and the spin-$\downarrow$ states is assumed to be weak and the characteristic
    length of its modulation is larger than the length scale of the trap,
    $\sqrt{\frac{\hbar}{m\omega}}$. Due to these assumptions it is reasonable to approximate the
    Rabi-coupling potential by its Taylor series.  We can, thus, demonstrate that our results
    generalize to all Rabi-coupling potentials with a non-vanishing second-order
    derivative\footnote{The Homogeneous term that is also contributing to the Taylor expansion of
    $H_S$, Eq. (6), preserves both of the $\hat S_z$ and $\hat S^2$ symmetries of $\hat{\tilde H}$
and consequently the its only effect is to shift the collective Larmor precession frequency.} by
showing that a similar dynamics as in section IIIA can be obtained for the parabolic spin-coupling
potential \begin{equation} \hat H'_S=\frac{1}{2}m(\Delta\omega)^2\int {\rm d}x~x^2
\sum_{\alpha,\beta}\hat \psi_\alpha^\dagger(x) \sigma^x_{\alpha \beta} \hat \psi_\beta(x).
\label{spin_dep_pot2} \end{equation} For our simulations we employ $\Delta\omega=\sqrt{0.1}$, while
the system is initialized in the fully polarized state, $| \Psi_F \rangle$, though as argued above a
similar dynamics takes place when the system is initialized in the $|\Psi_{S=3/2,Sz=1/2} \rangle$
state (results not shown here for brevity).  As it can be seen, the behaviour of the system in terms
of the spin-polarization, $P^{(1)}_S$ [see Fig. 6(c)] and the spin-spin correlations, $C^{(2)}_S$
[see Fig. 6(d)] is almost identical to the case of $\hat{H}_S$, Eq. (6) [see also Fig. 1(a) and
1(b)], with deviations occurring only within the $g \approx 0$ region [compare Fig. 6(c) and 1(b)].
We also note that in the case of strong spin-dependent potentials, where the exact shape of the
Rabi-coupling potential might play an important role, spin segregation phenomena are induced
\cite{AMR1,AMR2}. These compete with the ferromagnetic order identified here, as the overlap of the
spin-densities provides an upper bound for $\langle \hat S^2 \rangle$ and therefore such
investigations lie beyond the scope of this work.

\section{EXPERIMENTAL REALIZATION}

Our setup can be realized using $^{40}$K atoms under the influence of a Raman coupling of the two
energetically lowest hyperfine states, while the observables $P^{(1)}_S$ and $C^{(2)}_S$ are
accessible by fluorescence imaging.  Below we propose a possible experimental realization in order
to probe our findings. The robustness of the suggested measurement scheme is demonstrated by
comparing our MCTDHF results with simulated sets of single-shot images that contain additional noise
emulating this way the noise sources inherent in a corresponding experiment
\cite{Heidelberg,BrouzosJochim,ZurnAntiferro}.

\begin{figure*}[ht]
    \includegraphics[width=0.8\textwidth]{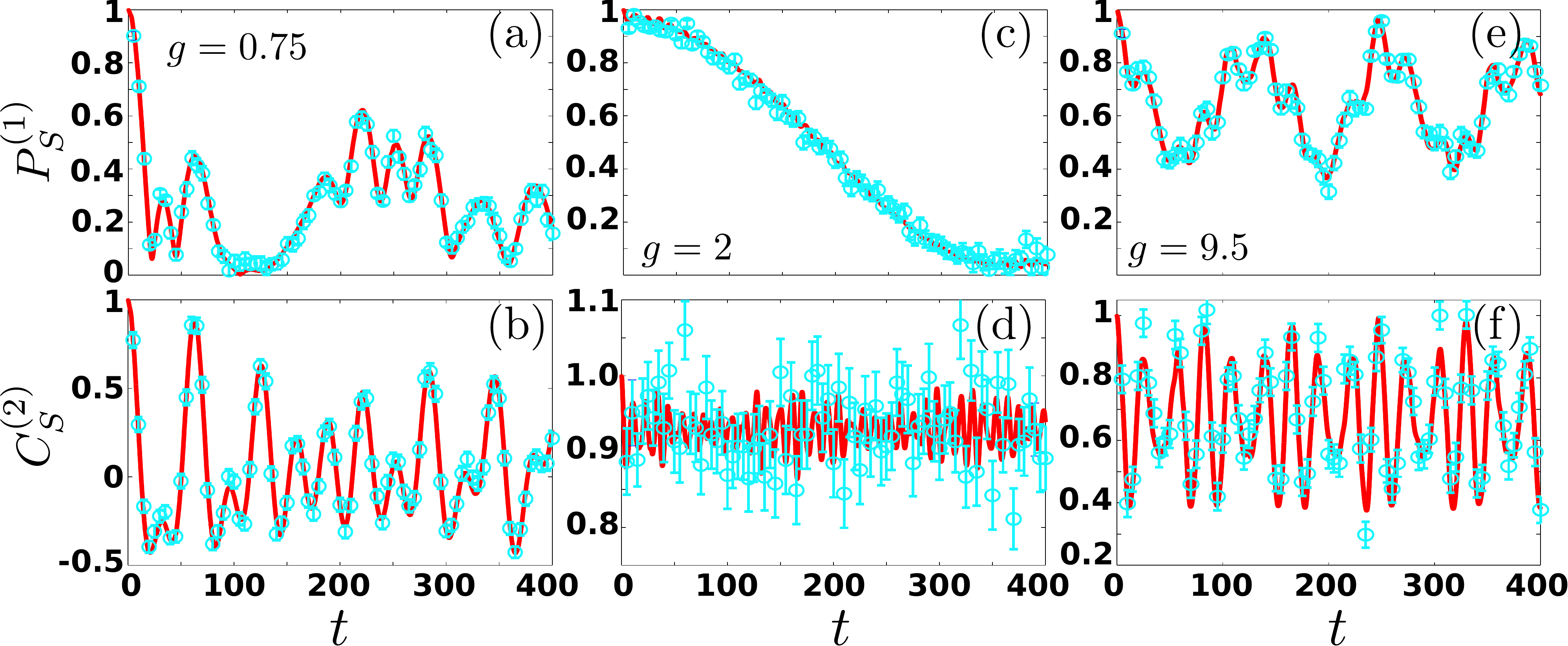}
    \caption{(a), (c), (e) Comparison of $P^{(1)}_S$
as calculated by MCTDHF (solid lines) and the average of $600$ single-shot measurements (data
points with error bars) within each of the dynamical regimes. For the single-shot simulations the
error $\Delta \phi=\pi/12$ has been incorporated. (b), (d), (f) Similar to (a), (c), (e) but for $C^{(2)}_S$. All quantities are given in harmonic units, error bars correspond to an
uncertainty of one standard deviation.}
    \label{fig:sys_sup}
\end{figure*}
The effective Rabi coupling scheme, see $\hat H_S$
Eq. (6), can be achieved by employing a two-photon resonant, $\delta=0$, Raman transition via two
Gaussian focussed laser beams.
To incorporate 
non-negligible interatomic interactions one needs to apply a bias magnetic field close to the point of an $s$-wave broad Fano-Feshbach
resonance \cite{Chin}.
For ${}^{40}$K atoms a broad $s$-wave Fano-Feshbach resonance between the hyperfine states $| \uparrow
\rangle = |{}^2S_{1/2};F=\frac{9}{2},m_F=-\frac{9}{2}\rangle$ and $| \downarrow \rangle
=|{}^2S_{1/2};F=\frac{9}{2},m_F=-\frac{7}{2}\rangle$, is located at the magnetic field
strength $B_{\rm FF}=202.10$ G \cite{Greinersup}. 

Fluorescence imaging is commonly used to probe the state of the system in few-atom ($N<10$) experiments
\cite{Heidelberg}. Here a certain number of atoms is ejected from the trap and recaptured into a
magneto-optical trap \cite{Pethicksup}. Subsequently, the number of ejected particles can be inferred by measuring the intensity of the
scattered light.
We show that $P^{(1)}_S$
and $C^{(2)}_S$ can be experimentally detected using fluorescence imaging.
$P^{(1)}_S$ and $C^{(2)}_S$ depend on the average and the variance of the
magnitude of the spin polarization respectively. Because of the employed Raman scheme the
Hamiltonian (6) is implemented in the interaction picture.
This implies that in the Schr{\" o}dinger picture and in the absence of the Raman fields the orientation of the spin-vector precesses around the $z$
spin-axis with frequency $\omega_{\uparrow \downarrow}=2\pi \times \left[ 44.8 + 0.156{\rm G}^{-1} ~(B -B_{\rm
FF})  \right]$MHz (where $B$ refers to the bias magnetic field). $\omega_{\uparrow \downarrow}$
    corresponds to the energy offset between the $| \uparrow \rangle$ and $| \downarrow \rangle$
    states of ${}^{40}$K for magnetic fields in the vicinity of $B_{\rm FF}$. As a consequence only the
spin-polarization along the $z$ axis (i.e. 
population-imbalance in the occupation of the hyperfine states $|\uparrow
\rangle$, $|\downarrow \rangle$) in spin-space can be directly probed.  To measure the
spin-state in such atomic systems Ramsay spectroscopy is employed to
coherently rotate the rotating $x$ or $y$ axes in the interaction picture to the stationary $z$ axis, which is common for both pictures.
A Ramsay spectroscopy sequence (described in the interaction picture) is utilized.
Initially all of the atoms are prepared in the
$N$-body state $|\Psi(0) \rangle$, namely all atoms reside in the $| \uparrow \rangle$ hyperfine
state. At time $t=0$ the
inhomogeneous Raman coupling  of the hyperfine
states is suddenly switched on and the fermions are exposed to it for time $t$.
By the end of this process, the MB wavefunction has evolved from $| \Psi(0) \rangle$ to $| \Psi(t)
\rangle$ (in the interaction picture) under the influence of $\hat H$.
At time $t$, the Raman coupling is suddenly switched off and the system
evolves for a dark time $t_{\rm dark}$. Within this time interval the reestablished
symmetries of the Hamiltonian $\hat{H}_0+\hat{H}_I$ [see Eq. (3), (4)]
prohibit any change to $\bm{S}$ and $S^2$. To measure the $S_x$ or $S_y$
components we need to rotate the desired spin component to the $z$ axis by
applying a $-\sigma_y$ or $\sigma_x$, $\frac{\pi}{2}$-pulse respectively by
means of spatially homogeneous two-photon-resonant optical Raman fields with the
appropriate phase shift, $\phi$, from the inhomogeneous one (for $S_z$ no $\frac{\pi}{2}$-pulse is used and the sequence continues
directly with the next step). This sequence stops the
precession dynamics of the desired spin component in the Schr{\" o}dinger picture
as it is mapped to the stationary $z$ axis. In the following, all
the spin-$\downarrow$ are removed from the trap by applying
a high-intensity resonant laser pulse at time $t_{\rm ex}$ \cite{ZurnAntiferro}. The surviving
atoms are loaded into the magneto-optical trap (at $t=t_{\rm meas}$) and counted to
provide a measurement for the spin polarization $S_i=\frac{2 N_\uparrow -N}{2}$
along the selected axis $i \in \{ x, y, z \}$ .

As a proof-of-principle of the above mentioned imaging procedure we simulate single experimental
measurements, where we take into account a random error in the phase $\Delta \phi$. We employ a generalized
version of the recent single-shot implementation offered by Multi-Layer Multi-Configuration Hartree method
for atomic Mixtures (ML-MCTDHX) \cite{MLX} (see \cite{kasparsup,Lodesup,filled_vortexsup,Lianew} and
appendix B for details) and evaluate $P_S^{(1)}$ and $C^{(2)}_S$ from the simulated experimental data.
Note that $\Delta \phi$ might be induced by variations in the optical path of the $-\sigma_y$ or $\sigma_x$,
$\frac{\pi}{2}$-pulse Raman beams.
To incorporate this source of error we simulate experimental measurements for each of the $x$, $y$ and $z$ components of the
spin vector and incorporate a random rotation by $\delta \phi$ along the $z$ axis that follows a Gaussian distribution of width $\Delta \phi$.

Figures 7(a) and 7(b) offer a comparison of our MCTDHF data with the simulated experimental estimates based on $600$
single-shot realizations containing an error $\Delta \phi= \pi/12$. We observe that
despite the latter error the single-shot results follow closely the MCTDHF data and reproduce both
the spin polarization $P_S^{(1)}$, as well as, the correlation dynamics $C^{(2)}_S$. The
uncertainty of the simulated single-shot results is of the order $\sim 0.05$. Agreement
between the MCTDHF data and the single-shot estimates is observed also for the ferromagnetically ordered
[see Fig. 7(c) and 7(d)] and the strong-$g$ demagnetization regime [see Fig. 7(e) and 7(f)]. Therefore, we
conclude that the error in the phase $\phi$ is not prohibitive for accurate measurements of $P^{(1)}_S$ and
$C^{(2)}_S$ as long as it is kept sufficiently smaller than $\frac{\pi}{2}$.

\section{SUMMARY AND OUTLOOK}
We have explored the spin-flip dynamics of few ultracold fermions subject to
spatially inhomogeneous external driving of the spins. In particular, we showed
that in this case the polarization of the confined Fermi gas cannot be
stabilized for any interaction strength. A result that lies in contrast to the
picture of ferromagnetism provided by the celebrated Stoner model.  Most
importantly, a stable correlation-induced ferromagnetic spin-order emerges in
spite of the strongly fluctuating polarization for moderate interactions. We
have characterized the emerging spin-order by comparing {\it ab initio}
simulations with an effective spin-chain model in the few-body case.  The
influence of correlations and the emergence of entangled NOON-like states
during the dynamical evolution of the system has been explicitly demonstrated.
In the weak and strong interaction limit the behavior of the system is
characterized by a significant depletion of the spin-spin correlator which can
be related to the corresponding avoided crossings appearing in the
eigenspectrum.  Our setup is experimentally accessible in $^{40}$K few-atom
experiments by employing a Raman coupling scheme of the two energetically
lowest hyperfine states. The observables $P_S^{(1)}$ and $C_S^{(1)}$ can be
measured by employing Ramsay spectroscopy and fluorescence imaging. 

It is known that the properties of itinerant magnetism vary depending on the confining
potential \cite{Zintchenko}. Studying the stability of ferromagnetism in the case of a double well potential or an optical lattice can
yield further insights into the magnetic properties exhibited in one dimensional systems.
Notice, also, that the spin-chain model presented here is easily extendable to higher dimensional
settings.
The investigation on whether a similar order
occurs in higher dimensional settings also provides an intriguing perspective for future study.
%yielding, also, an intriguing perspective for further studies.
Another interesting prospect is to examine the
demagnetization dynamics of few-fermions when exposed to Rashba and Dresselhaus spin-orbit coupling
\cite{Rashba,Dresselhaus}. This might establish a link to relevant condensed matter systems
where such demagnetization mechanisms are well-studied \cite{Dyakonov1,Elliot,Yafet}. Such dynamics
have recently been examined
in the case of thermal Fermi gases in the collisionless regime \cite{Radic,Natu,Stanescu}.

\appendix

\section{The Computational Method: ML-MCTDHX and the Spinor Variant of MCTDHF}

Our approach to solve the MB Schr\"{o}dinger equation $\left( {i\hbar {\partial _t} - \hat H}
\right) |\Psi (t) \rangle= 0$ relies on the 
Multi-layer Multi-Configuration Time-Dependent Hartree method for atomic Mixtures \cite{MLX}
(ML-MCTDHX). In particular we employ a reduction of the ML-MCTDHX method for spin-$1/2$ fermions
being referred in the following as the spinor-variant of Multi-Configuration
Time-Dependent Hartree method for Fermions (MCTDHF). 
MCTDHF has been applied extensively for the treatment of fermions with or without
spin-degrees of freedom, in a large class of condensed
matter, atomic and molecular physics scenarios (see e.g.
\cite{Zanghellinisup,Katosup,Caillatsup,Nestsup,Bonitzsup,Haxtonsup}) and recently also applied in the field of
ultracold atoms \cite{Axel,MLX,Jenny1,Jenny2,Fpolarons,CaoMistakidis}. The key idea of MCTDHF lies in the usage of a
time-dependent (t-d) and variationally optimized 
MB basis set, which allows for the optimal truncation of the total Hilbert space. 
The ansatz for the MB wavefunction is taken as a linear combination of t-d Slater
determinants
$|\vec{n} (t) \rangle$, with t-d weight coefficients $A_{\vec{n}}(t)$ 
\begin{equation}
    | \Psi (t) \rangle =\sum_{\vec{n}} A_{\vec{n}}(t) | \vec{n} (t) \rangle.
    \label{eq:mb_wfn}
\end{equation}
Each t-d Slater determinant is expanded in terms of $M$ t-d variationally optimized single-particle functions
(SPFs) 
$\left| \phi_l (t) \right\rangle$, $l=1,2,\dots,M$ with occupation numbers
$\vec{n}=(n_1,\dots,n_M)$. 
The SPFs are subsequently expanded in a
primitive basis $\lbrace \left| k,s \right\rangle \rbrace$, being the tensor product of a discrete
variable representation (DVR) basis
for the spatial degrees of freedom $\{ | k\rangle \}$ of dimension    
$M_{p}$ and the two-dimensional spin basis $\{ | \uparrow \rangle, | \downarrow \rangle \}$
\begin{equation}
    | \phi_j (t) \rangle= \sum_{k=1}^{M_p} \sum_{\alpha=\{ \uparrow, \downarrow \}} C^j_{k \alpha}
    (t) \ket{k} \ket{\alpha}.
    \label{eq:spfs}
\end{equation}
$C^j_{k \alpha}(t)$ refer to the corresponding t-d expansion coefficients.
Note
here that each t-d SPF is a general spinor wavefunction of the form $| \phi_j (t) \rangle= \int {\rm d}x~
[ \chi_j^\uparrow(x) \hat{\psi}^\dagger_\uparrow(x) + \chi_j^\downarrow(x)
\hat{\psi}^\dagger_\downarrow(x) ] | 0\rangle$ and hence the employed method is termed as the
spinor-variant of MCTDHF.
The time-evolution of the $N$-body wavefunction under the effect of the Hamiltonian $\hat{H}$
reduces to the determination of the $A$-vector coefficients and the SPFs, which in turn 
follow the variationally obtained MCTDHF equations of motion \cite{Alon,Axel,MLX}. 
In the limiting case of $M=N$, the method reduces to the t-d Hartree-Fock
method, while 
for the case of $M=2 M_{p}$, it is equivalent to a full configuration interaction approach
(commonly referred to as ``exact diagonalization'' in the literature)
 within the basis $\lbrace | k,s \rangle \rbrace$.

 For our implementation we have used a harmonic oscillator DVR, which results after
a unitary transformation of the commonly employed basis of harmonic oscillator eigenfunctions, as a
primitive basis for the spatial part of the SPFs.  To study the dynamics, we propagate the
wavefunction by utilizing the appropriate Hamiltonian within the MCTDHF equations of motion.  To
verify the accuracy of the numerical integration, we impose the following overlap criteria
$\left| \langle \Psi | \Psi \rangle - 1 \right| < 10^{-8}$ for the total wavefunction and $\left| \langle \varphi_i |
\varphi_j \rangle - \delta_{ij} \right| < 10^{-9}$  for the SPFs.
To infer about convergence, we increase the number of SPFs and DVR basis states such that the
observables of interest ($P_S^{(1)}$, $C_{s}^{(2)}$) do not change within a given relative accuracy
which is in our case $10^{-4}$. More specifically, we have used $M_{p}=60$, $M=26$ and $M_{p}=80$, $M=20$ for
the $N=3$ and the $N=5$ case respectively. Note that a full configuration interaction treatment
of the above-mentioned systems in the employed primitive bases would require $280840$ number-states for $N=3$ and $820384032$
ones for $N=5$.

\section{Single-Shot Procedure in Spin-$1/2$ Fermi Gases}\label{sec:single_shot}

The single-shot simulation procedure relies on a sampling of the MB probability distribution,
being available within the ML-MCTDHX framework.
In a spinor Fermi gas
the single-shot procedure is altered significantly when compared to the single component case \cite{kasparsup,Lodesup,filled_vortexsup}.
Here the role of entanglement between particles in different spin states plays
an important role.
For example consider the procedure that the spin-$\uparrow$ atoms are imaged before the
spin-$\downarrow$ atoms. Then, the total number of spin-$\uparrow$ atoms $N_\uparrow^{\rm imag}$ that will be imaged
is not {\it a priori} known due to the breaking of the $S_z$ symmetry. However, after imaging all of the
spin-$\uparrow$ atoms the number of spin-$\downarrow$ atoms is exactly known $N_\downarrow^{\rm
imag}=N-N_\uparrow^{\rm imag}$ since the total number of atoms $N$ is definite.

To capture the entanglement between the different spin states the MB wavefunction obtained
by ML-MCTDHX should be expressed such that the entanglement between the spin states is
evident. The spin-$1/2$ Fermi gas under consideration is a bipartite system \cite{Horodecki1sup,Horodecki2sup}
 since the
spatial degree of freedom for each particle in the spin-$\uparrow$ or spin-$\downarrow$
state resides in the Fock space $\mathcal{F}^{\uparrow}$, $\mathcal{F}^{\downarrow}$ respectively.
The latter results in a total Fock space $\mathcal{F}^{S=1/2}=\mathcal{F}^{\uparrow} \otimes\mathcal{F}^{\downarrow}$. 
Then, the MB wavefunction can be expressed in the Schmidt decomposition form (herewith we omit the
temporal dependence for simplicity)
\begin{equation}
    | \Psi \rangle= \sum_{k=1}^K \sqrt{\lambda_k} | \Psi_k^\uparrow \rangle | \Psi_k^\downarrow
    \rangle.
    \label{eq:Schmidt}
\end{equation}
The coefficient $\lambda_k$ is referred to as the natural occupation of the species
function $k$\footnote{The upper bound for summation reads
$K=\min(\dim(\mathcal{F}^{\uparrow}),\dim(\mathcal{F}^{\downarrow}))$ and it is therefore infinite in the general case even after considering that the $\mathcal{F}^\uparrow$ and $\mathcal{F}^\downarrow$
are restricted by the condition $N_\uparrow,N_\downarrow<N$. However, spinor MCTDHF truncates the
dimension of each Fock space to $\dim(\mathcal{F}^\alpha)=\sum_{N_{\alpha}=0}^N
\binom{M}{N_\alpha}$. For realistic applications even the latter value is too high and most of the
$\lambda_k$'s have a numerical-zero value. To cure this problem we
truncate the Schmidt decomposition further by setting $K=\min( K^{\rm eff} )$ obeying $\lambda_{k>K^{\rm eff}}<10^{-12}$.} Note that, $| \Psi_k^{\alpha} \rangle \in \mathcal{F}^{\alpha}$ and as
such the number of $\alpha$-spin particles varies for different Schmidt modes, $k$.
A state of the bipartite system [see Eq.
(\ref{eq:Schmidt})] cannot be expressed as a direct product of two states from the two subsystem
Fock spaces $\mathcal{F}^{\alpha}$ if at least two coefficients $\lambda_k$ are nonzero. In
the latter case the system is referred to as entangled \cite{Roncagliasup}.
The Schmidt decomposition of the MB wavefunction is obtained as follows. The reduced density
matrix for one of the spin states, let it be $\alpha$, is evaluated 
i.e. $\rho^{\alpha}=\mathrm{Tr}_{\alpha'} [ |\Psi
\rangle\langle \Psi | ]$, where $\alpha'$ refers to the spin state orthogonal to $\alpha$ and subsequently
diagonalized resulting in its Schmidt representation 
$\rho^{\alpha} = \sum_{k=1}^M \lambda_k |\Psi^{\alpha}_k \rangle \langle \Psi^{\alpha}_k|$. Then, the
corresponding species wavefunction of the spin state $\alpha'$ can be calculated by
$|\Psi_k^{\alpha'}
\rangle =\frac{1}{\sqrt{\lambda_k}} \langle \Psi_k^{\alpha} | \Psi \rangle$. 

The single-shot process in spinor gases represents a generalization of the single-shot process for a mixture with
a definite number of atoms in each species \cite{Lianew}. This generalization is based
on the treatment
of the vacuum state $|0^\alpha \rangle$. Before each step of the single-shot process the existence
of particles in the imaged spin state is checked. To perform the latter a random number in the interval
$P_{\rm rand} \in [0,1]$ is compared with $\lambda_{\tilde k}$, where $\tilde k$ is the Schmidt mode
for which $|
\Psi^\alpha_{\tilde k} \rangle =| 0^\alpha \rangle$ holds. If $P_{\rm rand} < \lambda_{\tilde k}$ the
imaging of the spin state $\alpha$ ends and the MB wavefunction is projected to 
$| \tilde \Psi \rangle = |0^\alpha \rangle \otimes | \Psi^{\alpha'}_{\tilde k} \rangle$.
Then the simulation of the imaging of the $\alpha'$ spin state is initiated. The MB wavefunction in
this case is the species wavefunction $| \Psi^{\alpha'}_{\tilde k} \rangle$ and as such the
single-shot procedure reduces to the well-established single species case (see
\cite{kasparsup,Lodesup,filled_vortexsup} and also the discussion below).
For $P_{\rm rand} > \lambda_{\tilde k}$ a particle in the spin state $\alpha$ is imaged.
First, a random position is drawn according to the constraint $\rho_{\alpha}^{(1)}(x_1')=\langle \Psi |
\hat{\psi}^\dagger_{\alpha}(x_1')\hat{\psi}_{\alpha}(x_1')| \Psi \rangle>l_1$ where $l_1$
refers to a random number within the interval [$0$, max\{$\rho^{(1)}_{\alpha}(x)\}$].
Then we project the $N$-body wavefunction to the ($N-1$)-body one by employing the
operator $\frac{1}{\mathcal{N}} \hat{\psi}_{\alpha}(x_1')$, where $\mathcal{N}=\sqrt{\langle \Psi | \hat
\psi^\dagger_{\alpha}(x_1') \hat \psi_{\alpha}(x_1') |\Psi \rangle}$ is a normalization factor.
The latter process directly affects the Schmidt coefficients $\lambda_k$'s (entanglement weights) and thus
despite the fact that the spin-$\alpha'$ atoms have not been imaged yet, both
$\rho_{\uparrow}^{(1)}(x)=\langle \Psi |
\hat{\psi}^{\dagger}_{\uparrow}(x)\hat{\psi}_{\uparrow}(x)| \Psi \rangle$ and $\rho_{\downarrow}^{(1)}(x)=\langle \Psi |
\hat{\psi}^{\dagger}_{\downarrow}(x)\hat{\psi}_{\downarrow}(x)| \Psi \rangle$ change.
This can be easily understood by employing again the Schmidt decomposition.
Indeed after this first measurement the ($N-1$)-particle MB wavefunction reads
\begin{equation}
        \ket{\tilde{\Psi}^{(-1)}}= \sum_k
        \sqrt{\tilde{\lambda}_k^{(-1)}}\ket{\tilde{\Psi}_{k}^{\alpha (-1)}}\ket{\Psi_k^{\alpha'}},
        \label{Eq:A1}
\end{equation}
where $\ket{\tilde{\Psi}_{k}^{\alpha
(-1)}}=\frac{1}{N_i}\hat{\psi}_{\alpha}(x_1')\ket{\Psi_k^{\alpha}}$ refer to the
 species wavefunction after the imaging  and
 $N_k=\sqrt{\bra{\Psi_k^{\alpha}}\hat{\psi}_{\alpha}^{\dagger}(x_1')\hat{\psi}_{\alpha}(x_1')\ket{\Psi_k^{\alpha}}}$
 denotes the corresponding normalization factor. Finally, the Schmidt coefficients read
 $\tilde{\lambda}^{(-1)}_{k}=\lambda_k N_k/\sum_m \lambda_m N_m^2$.
 The above-mentioned procedure is repeated $N^{\rm imag}_{\alpha}$ times until the condition $||\langle 0^{\alpha} |
 \Psi^{(-N^{\rm imag}_\alpha)} \rangle ||=1$ is reached or if a random number satisfying $P_{\rm rand} <
 \lambda_{\tilde k}^{(-N^{\rm imag}_\alpha)}$ is selected. The resulting distribution of
 positions ($x'_1$, \dots,$x'_{N^{\rm imag}_{\alpha}}$) is convoluted with a point spread function
 leading to a single-shot $\mathcal{A}^{\alpha}(\tilde{x})$ for the spatial configuration of
 spin-$\alpha$ particles,
where $\tilde{x}$ refers to the spatial coordinates within the image.
It is worth mentioning at this point that, in the special case for which the probability of $N^{\rm
imag}_{\alpha}=N$ is zero, it can be easily shown that upon annihilating the last spin-$\alpha$ particle (provided that
$P_{\rm rand} > \lambda^{(-N_{\alpha}^{\rm imag}+1)}_{\tilde k}$ is chosen) the $(N-N_\alpha^{\rm
imag})$-particle MB wavefunction becomes
\begin{equation}
        \ket{\tilde{\Psi}^{(-N^{\rm imag}_{\alpha})}}= \ket{0^\alpha} \otimes 
        \hspace{10pt}\sum_{k \neq \tilde{k}}
        \frac{\sqrt{\tilde{\lambda}^{(-N^{\rm imag}_{\alpha}+1)}_{k}} \braket{x'_{N_{\alpha}^{\rm imag}}|\Phi_{k}^{\alpha}}}
        {\sum\limits_{m \neq \tilde k} \sqrt{\tilde{\lambda}^{(-N^{\rm imag}_{\alpha}+1)}_{k}
        |\braket{x'_{N_{\alpha}^{\rm imag}}|\Phi_{k}^{\alpha}}|^2}}\ket{\Psi_k^{\alpha'}}.
        \label{Eq:A3}
\end{equation}
After this last step the entanglement between the spin states has been destroyed and the single
component wavefunction of
the spin $\alpha'$ atoms $\ket{\Psi^{(-N_{\alpha}^{\rm imag})}}$ corresponds to the second term on the right hand side of Eq. (\ref{Eq:A3}).

In this way, it becomes evident that after the imaging of spin $\alpha$ particles the resulting
wavefunction $\ket{\Psi^{(-N_{\alpha}^{\rm imag})}}=\ket{\Psi^{\alpha'}}$ [see Eq. (\ref{Eq:A3})] is a non-entangled ($N-N_{\alpha}^{\rm imag}$)-particle MB wavefunction and its corresponding single-shot procedure is
the same as in the single species case \cite{kasparsup}.
The latter is well-established (for details see \cite{kasparsup,Lodesup}) and here it is only
briefly outlined below.
We first calculate $\rho^{(1)}_{\alpha'}(x)$
from the MB wavefunction $\ket{\Psi^{(-N_{\alpha}^{\rm imag})}}=\ket{\Psi^{\alpha'}}$.
Then, a random position $x''_1$ is drawn obeying $\rho^{(1)}_{\alpha'}(x''_1)>l_2$ where $l_2$ is
a random number in the interval [$0$, $\max[\rho^{(1)}_{\alpha'}(x)]$].
Next, one particle located at position $x''_1$ is annihilated and $\rho^{(1) \prime}_{\alpha'}(x)$
is calculated from $\ket{\Psi^{(-N_{\alpha}^{\rm imag}-1)}}=\left[ \rho^{(1)}_{\alpha'}(x''_1) \right]^{-1/2} \hat \psi_{\alpha'}(x''_1)\ket{\Psi^{(-N_{\alpha}^{\rm imag})}}$.
To proceed, a new random position $x''_2$ is drawn from $\rho^{(1) \prime}_{\alpha'}(x;t_{im})$.
Following this procedure for $N-N_{\alpha}^{\rm imag}$ steps we obtain the distribution of positions
($x''_1$, \dots,$x''_{N-N_{\alpha}^{\rm imag}}$) which is then convoluted with a point spread function
resulting in a single-shot image $\mathcal{A}^{\alpha'}(\tilde{x}'|\mathcal{A}^{\alpha}(\tilde{x}))$.

\section{Spin-Chain Approach}

The spin-chain Hamiltonian builds upon the spin dependent
eigenstates, $| \chi_n^{\alpha} \rangle$, of the non-interacting Hamiltonian $\hat{H}_0 + \hat
H_S$, where $\alpha$, $n$ denote the spin and spatial modes. 
To simplify the notation below, we perform a rotation in spin-space by employing the
unitary operator $\hat{U}=e^{i \frac{\pi}{4} \hat{S}_y}$ such that the $x$ axis (see main text) is mapped to the
$z$ axis and thus the spin-modes correspond to $\alpha \in \{ \uparrow, \downarrow \}$.
The $n$-th spatial mode is considered
as singly occupied if either $| \chi_{n}^{\uparrow} \rangle$ or $| \chi_{n}^{\downarrow} \rangle$ is occupied, doubly occupied if both
are occupied and unoccupied if neither is occupied. Then the spatial
configurations are defined by $\vec{n}=(n_1,\dots,n_N)$ where $n_i$ refers to the occupied spatial
modes. There are $2^{N-2D}$ ($D$ denotes the number of double occupations) distinct states that correspond to the same spatial configuration, $\vec{n}$,
corresponding to the different available spin-configurations $\vec{\alpha}=(\alpha_1,\dots,\alpha_N)$. Consequently, a basis state of the $N$-body
system, $| \vec{n};\vec{\alpha} \rangle=\hat c^\dagger_{n_1 \alpha_1} \dots \hat c^\dagger_{n_N \alpha_N} |0\rangle$, is completely defined by its spin and
spatial configurations $\vec{\alpha}$ and $\vec{n}$ respectively.

To derive the effective Hamiltonian of the spin-chain model, $\hat H^{\rm eff}$, we
    neglect all terms that couple states of different spatial configurations.  The non-interacting
    Hamiltonian $\hat H_0 +\hat H_S$ is diagonal on the basis states $|\vec{n},\vec{\alpha}\rangle$
    and thus its exact form is incorporated in the effective spin-chain Hamiltonian, $\hat H^{\rm
    eff}$.  However, the same is not true for the interaction term $\hat H_I$. According to the
    above mentioned approximation, the general form of the effective interaction term, $\hat
    H_I^{\rm eff}$, contains all the terms in $\hat H_I$ that preserve the spatial configuration of
the state they act on.  There are only two terms in $\hat H_I$ that possess the latter property and
are linearly independent, namely the $\hat H_{I}^{\rm shift}$ and $\hat H_{I}^{\rm exc}$ terms.
$\hat H_{I}^{\rm shift}= g \sum_{m,n=0}^{m_{\rm max}} u_{mm}^{nn} \hat c^\dagger_{n\uparrow} \hat
c^\dagger_{m\downarrow} \hat c_{m\downarrow} \hat c_{n\uparrow}$ accounts for the energy shift of
the single-particle modes due to interaction, where $u^{nq}_{mp}=\int dx~ [ \chi^{\uparrow}_n(x)
\chi^{\downarrow}_m(x) ]^* \chi^{\downarrow}_p(x) \chi^{\uparrow}_q(x)$ denote the corresponding
interaction integrals.  $\hat H_I^{\rm exc} = g \sum_{m=0}^{m_{\rm max}} \sum_{n \neq m} u^{mn}_{nm}
\hat c^\dagger_{m\uparrow} \hat c^\dagger_{n\downarrow} \hat c_{m\downarrow} \hat c_{n\uparrow}$
allows for the exchange of the single-particle modes after a collision event.
Therefore, the effective Hamiltonian reads $\hat H^{\rm eff}=\hat H_0 +\hat H_S +
\hat H_I^{\rm eff}$, where $\hat H_I^{\rm eff} =\hat H_{I}^{\rm shift} + \hat H_{I}^{\rm exc}$.

To cast $\hat H^{\rm eff}$ in the spin-chain form we define the spin operators for each spatial mode 
        $\hat \sigma^0_n=\hat c_{n\uparrow}^\dagger \hat c_{n\uparrow} + \hat c_{n\downarrow}^\dagger \hat c_{n\downarrow}$, 
        $\hat \sigma^z_n=\hat c_{n\uparrow}^\dagger \hat c_{n\uparrow} - \hat c_{n\downarrow}^\dagger \hat c_{n\downarrow}$, 
        $\hat \sigma^+_n=\hat c_{n\uparrow}^\dagger \hat c_{n\downarrow}$ and 
        $\hat \sigma^-_n=\hat c_{n\downarrow}^\dagger \hat c_{n\uparrow}$.
The effective MB Hamiltonian $\hat{H}^{\rm eff}$
conserves the spatial modes for each spatial configuration $\vec{n}=(n_1,\dots,n_N)$ as it commutes with the projection operators 
\begin{equation}
    \hat{P}_{\vec{n}}=\sum_{\alpha_1=\{\uparrow,\downarrow\}} \dots
    \sum_{\alpha_N=\{\uparrow,\downarrow\}} \hspace{2pt}
    \hat c^\dagger_{n_1 \alpha_1} \dots \hat c^\dagger_{n_N\alpha_N}
    \hspace{2pt}|0\rangle \langle
    0| \hat c_{n_N\alpha_N} \dots \hat c_{n_1\alpha_1}.
    \label{eq:proj}
\end{equation}
Employing this projection operator, we can derive the spin-chain
Hamiltonian, $\hat
H_{\vec{n}}=\hat{P}_{\vec{n}} \hat{H}^{\rm eff} \hat{P}_{\vec{n}}$, for each configuration $\vec{n}$ with no double
occupations (i.e. $n_i \neq n_j$, $\forall~ i,j$), 
corresponding to the $N$-spin XXZ spin-chain
\begin{equation}
    \hat{H}_{\vec{n}}=\varepsilon_{\vec{n}}(g) + \sum_{i=1}^N h_{n_i}(g) \hat{\sigma}^z_{n_i}
    -g\sum_{i=1}^{N-1} \sum_{j=i+1}^N \bigg[\left( J^\perp_{n_i n_j} \hat{\sigma}^+_{n_i} \hat{\sigma}^-_{n_j}
    +J^{\perp *}_{n_i n_j} \hat{\sigma}^-_{n_i} \hat{\sigma}^+_{n_j} \right) 
    +J^z_{n_i n_j} \hat{\sigma}^z_{n_i} \hat{\sigma}^z_{n_j}\bigg].
    \label{eq:XXZ_1}
\end{equation}
The spin-spin interactions are given by the overlap integrals 
$J_{nm}^\perp=u^{mn}_{nm}$ and $J_{nm}^z=\frac{1}{4}(u^{nn}_{mm}+u_{nn}^{mm})$. The
interaction-dependent energy shift $\varepsilon_{\vec{n}}(g)$ and the local magnetic field
$h_{n_i}(g)$ read
\begin{equation}
    \begin{split}
    \varepsilon_{\vec{n}}(g)&=\sum_{i=1}^N \frac{E_{n_i}^\uparrow +E_{n_i}^\downarrow}{2}+g \sum_{i=1}^{N-1}
    \sum_{j=i+1}^N J^z_{n_i n_j},\\
    h_{n_i}(g)&=\frac{E_{n_i}^\uparrow -E_{n_i}^\downarrow}{2}+g 
    \sum_{j=1,~ j \neq i}^N \frac{u^{n_i n_i}_{n_j n_j}-u_{n_i n_i}^{n_j n_j}}{4}.
    \end{split}
    \label{eq:XXZ_terms}
\end{equation}

The configurations with $D$ double occupations have to be treated separately because the creation
operator of a double occupancy $\hat{c}^\dagger_{n\uparrow}\hat{c}^\dagger_{n\downarrow}$ possesses a
non-trivial commutation relation with the $\hat{\sigma}^0_n$ one, $[\hat{\sigma}^0_m,\hat{c}^\dagger_{n\uparrow}\hat{c}^\dagger_{n\downarrow}]=
2 \delta_{n m}\hat{c}^\dagger_{n\uparrow}\hat{c}^\dagger_{n\downarrow}$. In this case, it turns out
that the projected Hamiltonian, $\hat{H}_{\vec{n}}$, is expressed in terms of a $(N-2D)$-spin XXZ Hamiltonian,
$\hat{H}_{\vec{m}}^{(2D)}$,
with $\vec{m}$ being a $(N-2D)$-particle configuration composed of the singly occupied states of
$\vec{n}$. The $\hat{H}_{\vec{n}}$ and $\hat{H}_{\vec{m}}^{(2D)}$ are related via the creation operator of all the double occupations
$\hat{\mathcal{M}}^\dagger_{\vec{k}}\equiv \prod_{j=0}^{D-1} \hat{c}^\dagger_{k_j \downarrow}\hat{c}^\dagger_{k_j \uparrow}$,
where $\vec{k}$ is the vector of doubly occupied modes in $\vec{n}$, as $\hat{H}_{\vec{n}} = \hat{\mathcal{M}}^\dagger_{\vec{k}}
\hat{H}_{\vec{m}}^{(2D)} \hat{\mathcal{M}}_{\vec{k}}$. $\hat{H}_{\vec{m}}^{(2D)}$ has exactly the
same form as Eq. (\ref{eq:XXZ_1}) but the energy shift, $\varepsilon_{\vec{m}}^{(2D)}(g)$, and local
magnetic field, $h^{(2D)}_{m_i}(g)$, possess additional contributions when compared to the ones in
Eq. (\ref{eq:XXZ_terms}). Namely,
\begin{equation}
    \begin{split}
        \varepsilon_{\vec{m}}^{(2D)}(g)=&\varepsilon_{\vec{m}}(g)+ \sum_{i=1}^{D} 2 \varepsilon^0_{k_i}
         +g\Bigg[\sum_{i=1}^D 2 J^z_{k_ik_i}+ \sum_{i=1}^{D-1}
            \sum_{j=i+1}^{D} 4 J^z_{k_i k_j} +\sum_{i=1}^{D} \sum_{j=1}^{N-2D} J^z_{k_i m_j} \Bigg],\\
        h^{(2D)}_{m_i}(g)=&h_{m_i}(g) +g \sum_{j=1}^D \frac{u^{m_i m_i}_{k_jk_j}-u_{m_i m_i}^{k_jk_j}}{2}.
    \end{split}
    \label{eq:XXZ_2}
\end{equation}

The weight of each spatial configuration to the MB wavefunction $|\Psi(t)\rangle$ is
constant in time as the $w_{\vec{n}}=\langle \Psi(t) | \hat P_{\vec{n}} | \Psi(t) \rangle$
are conserved. Therefore, the time evolution of the MB wavefunction within the spin-chain
approximation reads
\begin{equation}
    |\Psi(t)\rangle=\sum_{\vec{n}} \sqrt{w_{\vec{n}}}~ e^{-i \hat H_{\vec{n}} t} |\Psi_{\vec{n}}(0) \rangle,
    \label{eq:mbwfn_sc}
\end{equation}
where $|\Psi_{\vec{n}}(0)\rangle= \frac{1}{\sqrt{w_{\vec{n}}}} \hat P_{\vec{n}} | \Psi (0) \rangle$
is the normalized initial wavefunction for each XXZ spin-chain. 

The generalization of the presented method compared to the one developed in Ref. \cite{AMR1} is the
inclusion of the interaction-dependent local magnetic potential [see Eq. (\ref{eq:XXZ_1}) and
(\ref{eq:XXZ_2})], which vanish for a linear gradient as the one considered in Ref. \cite{AMR1}
in the present case
such a term is important for obtaining the correct behaviour of the polarization magnitude $P^{(1)}_S$ in the ferromagnetically ordered regime. Within our implementation we numerically
diagonalize the one-body Hamiltonian, $\hat{H}_0+\hat H_S$, by employing the basis consisting of the $80$ energetically lowest
eigenstates of the harmonic oscillator and truncate the summation over $\vec{n}$ of Eq. (\ref{eq:mbwfn_sc}) by
taking into account only the contributions of spatial mode configurations with $|w_{\vec{n}}|>
10^{-12}$. This truncation results in $1520$ and $38304$ configurations for $N=3$ and $N=5$ fermions respectively.

\begin{acknowledgements}

This work is supported by the Cluster of Excellence
`Advanced Imaging of Matter' of the Deutsche Forschungsgemeinschaft (DFG) - EXC 2056 - project ID 390715994.
S.I.M. and P.S. gratefully
acknowledge financial support by the Deutsche Forschungsgemeinschaft (DFG) in the framework of
the SFB 925 ``Light induced dynamics and control of correlated quantum systems''.
\end{acknowledgements}
{}

\end{document}